\documentclass[12pt]{article}
\usepackage{amssymb}
\usepackage{amsmath}

\usepackage{multirow} 
\usepackage{amstext}
\usepackage{graphicx,epsfig}
\usepackage{epsfig}
\usepackage{verbatim} 
\usepackage{fancyhdr}
\usepackage{fancybox}
\usepackage{color}
\usepackage{ulem,bbold}
\usepackage{enumitem}
\usepackage{subfigure}
\usepackage{bbm}
\usepackage{parskip}
\usepackage{ytableau}
\usepackage{longtable}

\usepackage{array}
\usepackage{tensor}
\usepackage{tikz}
\usepackage{cite}
\usepackage{cite}

\newcommand{\Comment}[1]{{}}
\definecolor{MyDarkBlue}{rgb}{0.15,0.15,0.45}
\usepackage[linktocpage=true]{hyperref}
\hypersetup{
colorlinks=true,
citecolor=MyDarkBlue,
linkcolor=MyDarkBlue,
urlcolor=MyDarkBlue,
}

\usepackage{genyoungtabtikz}
\Yboxdim{13pt} 
\Ylinethick{.8pt}

\newcommand{\be}{\begin{equation}}
\newcommand{\ee}{\end{equation}}
\newcommand{\bea}{\begin{eqnarray}}
\newcommand{\eea}{\end{eqnarray}}
\newcommand{\beas}{\begin{eqnarray*}}
\newcommand{\eeas}{\end{eqnarray*}}
\newcommand{\nn}{\nonumber}

\numberwithin{equation}{section}

\begin{document}

\title{).}

\begin{center}
{\Large \bf{Conformal Killing Tensors, Noether Currents and \\ \vspace{.1cm} Higher-Spin Shift Symmetries in (A)dS Space}}
\vspace{.2cm}
{\Large \bf{  }}
\end{center}

\vspace{1truecm}
\thispagestyle{empty}
\centerline{\Large Kurt Hinterbichler,${}^{\rm }$\footnote{\href{mailto:kurt.hinterbichler@case.edu} {\texttt{kurt.hinterbichler@case.edu}}} Samanta Saha,${}^{\rm }$\footnote{\href{mailto:sxs2638@case.edu} {\texttt{sxs2638@case.edu}}} Thomas Yan${}^{\rm }$\footnote{\href{mailto:txy209@case.edu}{\texttt{txy209@case.edu}}}  }

\vspace{.5cm}

\centerline{{\it ${}^{\rm }$CERCA, Department of Physics,}}
\centerline{{\it Case Western Reserve University, 10900 Euclid Ave, Cleveland, OH 44106}} 
\vspace{.25cm}

\begin{abstract} 

For certain mass values, shift symmetries appear among massive higher spin fields propagating on (anti-) de Sitter spacetime.  On the one hand, Noether’s theorem assigns a set of conserved currents for each shift symmetric field, one current for each of the independent shift symmetries.  On the other hand, each shift symmetric field comes with a higher-rank conserved field strength that can be contracted with a conformal Killing tensor (CKT) to form another set of conserved currents, one for each independent CKT.  This second set is naively much larger than the first.  We conjecture, and prove in the first few cases, that these two sets are the same once we account for the redundancy due to trivial currents that is implicit in Noether's theorem.  For each field, only one branch of the CKTs is non-trivial.   As we range over all the mass values and spins of the shift symmetric fields, each kind of CKT gets used exactly once in a non-trivial current.

\end{abstract}

\newpage

\thispagestyle{empty}

\tableofcontents

\setcounter{page}{1}
\setcounter{footnote}{0}

\parskip=5pt
\normalsize

\section{Introduction\label{introsec}}

Symmetry principles play a central role in constraining the dynamics of field theories. 
A pillar of our understanding of symmetry in this context is Noether’s first theorem~\cite{Noether:1918zz}, which connects continuous symmetries to conserved quantities: 
for every continuous symmetry of the action, there exists a current that is conserved on shell.

The symmetries we will be interested in are extended shift symmetries among massive higher spin fields in (anti-)de Sitter ((A)dS) space~\cite{Bonifacio:2018zex}.  These are nonlinearly realized (i.e. spontaneously broken) spacetime symmetries, 
generalizing to (A)dS and to higher spins the simple shift symmetry of a massless scalar field, as well as the extended versions such as the Dirac-Born-Infeld and special galileon symmetries that underlie the known exceptional effective field theories ~\cite{Goon:2011qf,Goon:2011uw,Burrage:2011bt,Cheung:2014dqa,Hinterbichler:2015pqa,Cheung:2016drk,Novotny:2016jkh,Padilla:2016mno,Bogers:2018zeg,DeRham:2018axr,Bonifacio:2019hrj,Bonifacio:2021mrf}.

A massive bosonic field of spin $s$ on $D$ dimensional (A)dS space is described by a completely symmetric tensor $\phi_{\mu_1 \cdots \mu_s}$ obeying the on shell equations of motion
\bea
&& \begin{cases}  (\Box - m^{2})\,\phi = 0 \, , & s = 0  \,, \nn\\ 
\left( \Box - H^{2}\bigl[s + D - 2 - (s-1)(s + D - 4)\bigr] - m^{2} \right)
\phi_{\mu_{1}\cdots\mu_{s}} = 0 , & s \ge 1\, ,  
\end{cases} \nn\\
&&\nabla^{\nu}\phi_{\nu\mu_{2}\cdots\mu_{s}} = 0 \, , \nn\\
&&\phi^{\nu}{}_{\nu\mu_{3}\cdots\mu_{s}} = 0 \,. \label{eqnofmotion}
\eea

(Here and throughout we will use the signs appropriate for dS space, where $1/H$ is the dS radius.  Expressions for AdS can be obtained by taking $H^2\rightarrow 1/L^2$ with $L$ the AdS radius.)
As discussed in \cite{Bonifacio:2018zex}, these theories become shift-symmetric at the following particular mass values,
\begin{equation}
\left\{
\begin{aligned}
m_k^{\,2} &= -k(k + D - 1)H^2, && s = 0\, , \\[6pt]
m_k^{\,2}  &= -(k + 2)\bigl(k + D - 3 + 2s\bigr)H^2, && s \ge 1\,.
\end{aligned}
\right.
\qquad k = 0,1,2,\ldots \label{shifssmgeee}
\end{equation}
The integer $k$ is called the level of the shift symmetry. The explicit form of the shift symmetry is given by
\begin{equation}
\delta \phi_{\mu_1 \cdots \mu_s}
= S_{A_1 \cdots A_{s+k},\, B_1 \cdots B_s}\;
X^{A_1} \cdots X^{A_{s+k}} \nabla_{\mu_1} X^{B_1}\cdots \nabla_{\mu_s} X^{B_s}\, ,
\label{shifttransformation}
\end{equation}
where $X^A(x)$ is the standard embedding of dS$_D$ into an auxiliary flat Lorentzian ambient space of dimension $D+1$ with metric $\eta_{AB} = \mathrm{diag}(-1, 1, \ldots, 1)$ (for details and for our conventions of the embedding formalism see Appendix~A of~\cite{Bonifacio:2018zex}). The tensor $S_{A_1 \cdots A_{s+k},\, B_1 \cdots B_s}$ is a constant embedding space tensor with the index symmetries of the following completely traceless two-row Young tableau,
\be
S_{A_1\cdots A_{s+k},\, B_1\cdots B_s} \in  ~\raisebox{1.15ex}{\gyoung(_5{s+k},_3{s})}\,.
\ee
(We use the anti-symmetric convention for tableaux.)

On dS$_D$, in the scalar case and when the shift symmetries are gauged, the shift symmetric fields describe unitary irreducible representations of the dS$_D$ isometry group, known as the discrete series representations in $D=2$ and the exceptional series representations in higher dimensions~\cite{Dobrev:1977qv,Boers:2013pba,Basile:2016aen,Sun:2021thf,Sengor:2022lyv,Sengor:2022kji,Enayati:2022hed,RiosFukelman:2023mgq,Anninos:2023lin,Schaub:2024rnl,Chen:2026kjo}.  If the shift symmetry is not gauged, they are non-unitary.  For the higher spins $s>0$, they are always non-unitary (except for the vector in $D=2$, which is dual to the scalar, when both are gauged \cite{Hinterbichler:2024vyv,Farnsworth:2024yeh}).  On AdS$_D$, on the other hand, the shift symmetric fields are always unitary, whether the symmetry is gauged or not, and it is conjectured that they play a role in some supergravity compactifications and holographic setups \cite{Lee:1999yu,Alkalaev:2019xuv,Conlon:2021cjk,Apers:2022tfm,Apers:2022vfp,Plauschinn:2022ztd,Blauvelt:2022wwa,Arboleya:2025ocb,Bekaert:2025azj}.

We will be interested in the case where the shift symmetry is not gauged: we consider \eqref{shifttransformation} as a global symmetry and look for the Noether currents.  Given an action depending only on a set of fields $\phi^I$ and their first derivatives $\nabla_\mu \phi^I$, 
\be S=\int d^Dx\sqrt{|g|}{\cal L}\left(\phi,\nabla \phi\right)\,,\label{actionwe}\ee
a continuous symmetry is any infinitesimal transformation $\delta \phi^I$ that leaves the Lagrangian ${\cal L}$ as defined in \eqref{actionwe}, i.e. without the $\sqrt{|g|}$ factor, invariant up to a total covariant derivative,\footnote{We will be working on a fixed de Sitter background so we will use covariant conservation.  Noether's theorem is usually expressed in terms of ordinary conservation, but on a fixed background there is no essential difference since $\sqrt{|g|}\nabla_\mu {\cal J}^\mu=\partial_\mu\left(\sqrt{|g|} {\cal J}^\mu\right)$.}
\be\delta{\cal L}= \frac{\partial \mathcal{L}}{\partial \phi^I} \, \delta \phi^I 
  + \frac{\partial \mathcal{L}}{\partial (\nabla_\mu \phi^I)} \, \nabla_\mu \delta \phi^I =  \nabla_\mu {\cal K}^\mu\,,\label{dlisgdke}\ee
where ${\cal K}^\mu$ can depend on the field and any of its higher derivatives.  Given this, it is straightforward to see that the current
\be \mathcal{J}^\mu = \frac{\partial \mathcal{L}}{\partial (\nabla_\mu \phi^I)} \delta\phi^I - {\cal K}^\mu \,,\label{exprnere}\ee
is covariantly conserved on the equations of motion,
\be \nabla_\mu \mathcal{J}^\mu = 0 \ \ \ {\rm when} \ \  \ \frac{\partial \mathcal{L}}{\partial \phi^I}
    - \nabla_\mu \frac{\partial \mathcal{L}}{\partial (\nabla_\mu \phi^I)}=0\,.
\ee

There is an inherent ambiguity in the definition of ${\cal J}^\mu$ given by \eqref{exprnere} that will be important for what follows: the variation of the Lagrangian in \eqref{dlisgdke} gives us only the divergence of ${\cal K}^\mu$, and we must strip off the derivative to get ${\cal K}^\mu$ by itself in order to write the current in \eqref{exprnere}.   Stripping off the derivative leaves ${\cal K}^\mu$, and hence the current given by \eqref{exprnere}, ambiguous up to the addition of any term which is itself a divergence, as follows:
\be {\cal J}^\mu \sim {\cal J}^\mu +\nabla_\nu {\cal I}^{\mu\nu}\, .\label{improgee}\ee
Here ${\cal I}^{\mu\nu}$ is any tensor, which can depend on the field and any of its higher derivatives, which is anti-symmetric on shell.   This additional term in \eqref{improgee}, usually called an improvement term, does not spoil the conservation of ${\cal J}^\mu$ on shell, since $\nabla_\mu \nabla_\nu {\cal I}^{\mu\nu}$ vanishes identically on shell.  Conserved currents should thus be thought of as belonging to equivalence classes, with two currents considered equivalent if they differ by an improvement term.  Noether's theorem assigns only an equivalence class, not a unique current, to each symmetry \cite{Brandt:1989gy,Wald:1990mme,Dubois-Violette:1991dyw,Barnich:2000zw,Barnich:2001jy}.

For the shift symmetric fields, we thus expect an equivalence class of Noether currents for each shift symmetry \eqref{shifttransformation}, i.e. one for each independent value of the mixed symmetry tensor $S_{A_1 \cdots A_{s+k},\, B_1 \cdots B_s}$ parametrizing the transformations.

There is, however, another natural set of conserved currents that comes more directly out of the construction of the shift symmetric fields.   
There is a basic ``field strength''  operator in the theory \cite{Bonifacio:2018zex}: it is a symmetric and traceless tensor $F_{\mu_1  \cdots \mu_{s+k+1}}$ of rank $s+k+1$, which is constructed by acting with $k+1$ derivatives on the field and then taking the symmetric traceless part,
\begin{equation}
F_{\mu_1 \cdots \mu_{s+k+1}}
= \nabla_{(\mu_{s+1}} \cdots 
  \nabla_{\mu_{s+k+1}} 
  \phi_{\mu_1 \cdots \mu_s)_T} \, .\label{fieldstrnree}
\end{equation}
(Here $(\ldots)_T$ indicates the fully symmetric and fully traceless part of the enclosed indices.)
The field strength is divergenceless on shell,
\be \nabla^{\mu_1} F_{\mu_1 \mu_2 \cdots \mu_{s+k+1}} = 0 \,,\label{consonheffee}\ee 
as one can directly check from the equations of motion~\eqref{eqnofmotion}, and it is invariant under the shift symmetries \eqref{shifttransformation}.
Given a divergenceless higher spin operator such as this, a standard conserved current can be constructed from it by contracting all but one of its indices with a conformal Killing tensor (CKT).  A rank $r$ CKT on dS$_D$ is a symmetric traceless tensor $\xi_{\mu_1\cdots \mu_r}$ that satisfies the conformal Killing equation,
\be \nabla_{(\mu_{1}} \xi_{\mu_2\cdots \mu_{r+1})_T}=0\, . \label{conforkee}\ee
For each such CKT of rank $r=s+k$, we can contract it with the field strength to form a current,
\be  J^\mu = F^{\mu\mu_1 \cdots \mu_{s+k}} \xi_{\mu_1\cdots  \mu_{s+k}}\,,\label{fjojccree}\ee
which will be conserved on shell by virtue of \eqref{consonheffee}, \eqref{conforkee}, and the symmetry and tracelessness of the field strength.\footnote{The most familiar example of the contraction of a higher-rank conserved operator with a Killing object in order to make a conserved current occurs in conformal field theory, where the symmetric and traceless conserved stress tensor $T_{\mu\nu}$ is contracted with the ordinary conformal Killing vectors $\xi_\mu$ satisfying $ \nabla_{(\mu} \xi_{\nu)_T} =0$ to make the conserved currents $J^\mu\sim T^{\mu\nu}\xi_\nu$, whose charges generate the conformal symmetries of the theory.  Our construction here for shift symmetries, using higher rank operators and Killing tensors, mirrors that for higher spin symmetries in CFT \cite{Mikhailov:2002bp,Maldacena:2011jn}, the difference is that in our case the shift symmetries are non-linearly realized, so the conserved operators start linearly in the fields rather than quadratically. (See \cite{Antunes:2025iaw} for recent constraints on linearly realized higher spin symmetries of massive fields on AdS.)}

The space of solutions to \eqref{conforkee} is finite dimensional \cite{Eastwood:2002su,dairbekov2011conformalkillingsymmetrictensor}, so there are a finite number of CKTs and hence a finite number of conserved currents that one can construct from the field strength as in \eqref{fjojccree}.  As we will see, except for the case of the simple shift symmetric scalar at $k=0$, this space of solutions is always larger than the space of Noether currents coming from the shift symmetries.  
This apparent mismatch will be resolved by taking into account the redundancy \eqref{improgee} up to improvement terms present in the currents.  We will see that only one class of solutions to the conformal Killing equations gives rise to non-trivial conserved currents, and it is these that correspond to the Noether currents of the shift symmetry.

The simplest example that illustrates this is the $k=1$ shift-symmetric scalar field with mass $m^2 = - D H^2$.  The conserved field strength is the rank 2 symmetric traceless tensor 
\be F_{\mu\nu}=\nabla_{(\mu}\nabla_{\nu)_T}\phi\,,\label{Finfdke1edefee}\ee
and from this
 one can construct the conserved currents 
 \be J^\mu = F^{\mu\nu}\xi_\nu\, ,\ee 
 where $\xi^\mu$ is any conformal Killing vector on dS$_D$, satisfying $\nabla_{(\mu} \xi_{\nu)_T} =0$.
On dS$_D$, there are $(D+1)(D+2)/2$ linearly independent conformal Killing vectors, and they naturally split into two different branches.  One branch consists of the Killing vectors of dS space; these are the conformal Killing vectors $\xi^\mu_{(1)}$ which are also divergenceless, $\nabla_\mu\xi^\mu_{(1)}=0$, so that they satisfy the Killing equation $\nabla^{(\mu}\xi_{(1)}^{\nu)} = 0$.
In terms of ambient space coordinates, they can be written as
\be
\xi^{\mu}_{(1)} = S_{A,B} X^{A} \nabla^{\mu} X^{B}\,,
\ee
where 
\be S_{A,B}\in ~\raisebox{1.15ex}{\yng(1,1)}   \ee
is a constant antisymmetric tensor in the ambient space, parametrizing the $D(D+1)/2$ conformal Killing vectors of this type. The other branch consists of the conformal Killing vectors ${\xi}^\mu_{(0)}$ that are gradients of a scalar, i.e. the genuine conformal Killing vectors that are not Killing vectors.  These can be written as
\be
{\xi}^\mu_{(0)} = \nabla^\mu (S_A X^A)\,,
\ee
where
\be S_A \in ~\raisebox{-0.15ex}{\yng(1)}  \ee
is a constant ambient-space vector, so that there are $D+1$ conformal Killing vectors of this type.  From the conformal Killing vectors in these two branches, we can construct two sets of conserved currents using \eqref{fjojccree},
\be
J^\mu_{(1)} = F^{\mu\nu}\xi_{(1)\nu}\, ,
\qquad
{J}^{\mu}_{(0)} =F^{\mu\nu}\xi_{(0)\nu}\, .
\ee

On the other hand, the $k=1$ shift transformation reads
\be \delta \phi = S_AX^A\,,\ee
parametrized by the ambient space vector $S_A$, and
the Noether procedure thus yields a set of $D+1$ conserved currents from these symmetries, which are
\be {\cal J}^\mu = \phi \nabla^\mu\left(S_AX^A\right)-{\nabla^\mu} \phi\left(S_AX^A\right)\,  . \label{noetherk1e}\ee
These correspond in their number to ${J}^{\mu}_{(0)}$, but they have fewer derivatives acting on the field.  

There are no Noether currents corresponding to the other branch of currents, $J^\mu_{(1)}$. This prompts the question of whether the Killing vectors realize additional hidden global symmetries in (A)dS, beyond the shift symmetry itself. In this example, we can see that they do not, and that they are instead trivial currents that can be written as a pure improvement term: using \eqref{Finfdke1edefee}, the Klein Gordon equations $(\Box+DH^2)\phi=0$ and $(\Box+(D-1)H^2)\xi^\mu_{(1)}=0$ satisfied by the field and the Killing vectors, as well as transversality of the Killing vector $\nabla_\mu \xi^\mu_{(1)}=0$, we can write
\begin{equation}
   J_{(1)}^\mu = F^{\mu\nu}\xi_{(1)\nu} =\nabla_\nu\left( 2 \nabla^{[\mu}\phi\xi_{(1)}^{\nu]} -2 \phi\nabla^{[\mu}\xi_{(1)}^{\nu]}  \right) \,,
\end{equation}
demonstrating that $J^\mu_{(1)}$ can be reduced to pure improvement terms and thus is equivalent to a trivial current under the equivalence \eqref{improgee}. 

For the other branch, we can write
\begin{equation}
   {J}^{\mu}_{(0)} = F^{\mu\nu}{\xi}_{\nu(0)}=\nabla_\nu\left(2\nabla^{[\mu}\phi\nabla^{\nu]}(S_A X^A)\right)-H^2(D-1)(\phi\nabla^\mu(S_A X^A)-\nabla^\mu\phi(S_A X^A))\,.
 \end{equation}
The first term on the right-hand side is an improvement term, and the second term is proportional to \eqref{noetherk1e},
demonstrating that ${J}^{\mu}_{(0)}$ is proportional to the Noether current up to improvement.  

Thus only one of the two branches of conformal Killing tensors ends up giving a non-trivial current, and this non-trivial current corresponds to the Noether current of the shift symmetry.

In what follows, we will detail how this works for all the spins and values of $k$: in each case, there is one branch of conformal Killing tensors that corresponds to the Noether currents of the shift symmetries, and the remaining branches are trivial.  Though it appears that many conformal Killing tensors are thus wasted, we will see that as we range over all the values of $s$ and $k$, each kind of conformal Killing tensor gets used non-trivially exactly once, and so there is in the end a correspondence between the shift symmetries among all the shift symmetric fields and the full set of conformal Killing tensors.  This correspondence is illustrated in figure \ref{crossingfig}. We do not yet have a full proof of these statements; in section \eqref{examplessec} we will check explicitly that they are true for the lowest few values of $s$ and $k$ and leave the remainder as a conjecture.

\section{Conformal Killing Tensors on dS$_D$\label{CKTsection}}

We start with the classification of CKTs on dS$_D$, which will be contracted with the field strengths \eqref{fieldstrnree} to form the CKT currents \eqref{fjojccree}.  A rank $r$ CKT is a completely symmetric and traceless tensor $\xi_{\mu_1\cdots \mu_{r}}$ that solves the conformal Killing equation,
\be \nabla_{(\mu_{1}} \xi_{\mu_2\cdots \mu_{r+1})_T}=0\, . \label{conforkee2}\ee
This equation is Weyl invariant: the solutions are the same for any two metrics related by a Weyl transformation (where the CKT with upper indices, $\xi^{\mu_1\cdots \mu_{r}}$, is what does not transform) \cite{HEIL2016383}, and so the classification of solutions is the same as that of flat space \cite{Eastwood:2002su}.

The general global CKTs are constructed by taking gradients of a set of tensor spherical harmonics on dS$_D$, which we call modes.  The set of rank $r$ CKTs splits into $r+1$ branches; each branch is labelled by an integer $l$ which indicates the mode it will be constructed from, and which ranges over
\be l=0,1,2,\ldots,r.\ee
The $l$-th mode of the rank $r$ Killing tensor is the rank $l$ tensor spherical harmonic $\zeta^{(l,r)}_{\mu_1\cdots \mu_{l}}$, given in terms of the embedding space by 
\be \zeta^{(l,r)}_{\mu_1\cdots \mu_{l}}= {S}_{A_1\cdots A_r,B_1\cdots B_l}X^{A_1}\cdots X^{A_{r}}\nabla_{\mu_1}X^{B_{1}}\cdots\nabla_{\mu_l}X^{B_{l}}\,,\label{modesdefe}\ee
where $S_{A_1\cdots A_r,B_1\cdots B_l}$ is a constant embedding space tensor which is fully traceless and has the symmetries of the Young tableau
\be S_{A_1\cdots A_r,B_1\cdots B_l} \in ~\raisebox{1.15ex}{\gyoung(_5{r},_3{l})}\,.\label{Cyoungseyme} \ee
The modes \eqref{modesdefe} are symmetric and traceless (as can be seen from the identity $\nabla_{\mu}X^{A}\nabla^{\mu}X^{B}=\eta^{AB}-H^2X^AX^B$ and the Young symmetry \eqref{Cyoungseyme}), they satisfy a Klein-Gordon equation \cite{BRANSON1992314},
\begin{equation}
    \left[\square+H^2\left(r(r+D-1)-l \right)\right]\zeta^{(l,r)}_{\mu_1\cdots \mu_{l}}= 0\,,\label{modekhgeefe}
\end{equation}
and they are divergenceless,
\be \nabla^{\mu_1} \zeta^{(l,r)}_{\mu_1\mu_2\cdots \mu_{l}}=0\,.\label{modeedivergee}\ee

The CKTs are constructed from the modes by taking symmetrized and traceless covariant derivatives as follows,
\begin{equation}
   \xi^{(l,r)}_{\mu_1\cdots \mu_{r}}=\nabla_{(\mu_1}\cdots\nabla_{\mu_{r-l}}\zeta^{(l,r)}_{\mu_{r-l+1}\cdots\mu_r)_T}\,. \label{symconvee}
\end{equation}
The CKT is labelled by its rank $r$, and by the rank $l$ of its mode, and has $r-l$ derivatives acting on the mode.  
Note that the case $l=r$ is where no derivatives are being taken in \eqref{symconvee}, and so the mode in this case is itself a CKT, $\xi^{(r,r)}=\zeta^{(r,r)}$.  The CKTs satisfy a Klein-Gordon equation with a mass that depends on the mode $l$ and the rank $r$.  Using \eqref{modekhgeefe} and Eq. 2.40 of \cite{Hinterbichler:2024vyv} we have
\begin{equation}
   \left[\Box+H^2\left(l (l+D-3) + r\right)\right]\xi^{(l,r)}_{\mu_1\cdots\mu_r}=0\,.
\end{equation}
Those with $l\geq 1$ also satisfy a multiple divergence condition,
\be  \nabla^{\mu_1}\cdots\nabla^{\mu_{r-l+1}} \xi^{(l,r)}_{\mu_1\cdots \mu_{r}}=0 \,,\ \ l=1,2,\ldots, r\,.\ee
In particular, when $l=r$, where we have $\xi^{(r,r)}=\zeta^{(r,r)}$, they are divergenceless as in \eqref{modeedivergee}: these are the Killing tensors, satisfying $\nabla_{(\mu_{1}} \xi^{(r,r)}_{\mu_2\cdots \mu_{r+1})}=0$ \cite{Barnich:2005bn}.  The modes $l<r$ then parametrize the CKTs that are not Killing tensors.

For $r=0$ we have $l=0$ and the only CKT is a constant scalar.  For $r=1$, we have the case discussed in the introduction: $l=1$ gives the Killing vectors, parametrized by $S_{A,B}\in ~\raisebox{1.15ex}{\yng(1,1)}\ $, and $l=0$ gives the conformal Killing vectors which are not Killing vectors, formed from the gradients of scalar modes parametrized by $S_A\in ~\raisebox{0.0ex}{\yng(1)}\ $.    

For a general rank $r$ CKT, its modes are parametrized by tensors with the following tableaux:
\be \overset{l=r}{ ~\raisebox{1.15ex}{\gyoung(_5{r},_5{r})}}\,, \  \overset{l=r-1}{ ~\raisebox{1.15ex}{\gyoung(_5{r},_4{r-1})} }\,, \  \overset{l=r-2}{ ~\raisebox{1.15ex}{\gyoung(_5{r},_3{r-2})}}\, , \ \cdots \ \, , \  \overset{l=1}{  ~\raisebox{1.15ex}{\gyoung(_5{r},_1)}}\, , \   \overset{l=0}{ ~\raisebox{1.15ex}{\gyoung(_5{r})} } \,.\label{conformreop}\ee
Note that these are precisely the representations that would be obtained upon dimensional reduction of a single traceless tensor of type ${~\raisebox{1.15ex}{\gyoung(_5{r},_5{r})}}$ living in a $D+2$ dimensional space.  This is because all the rank $r$ CKTs taken together form an irreducible (for generic D) representation of the conformal algebra of dS$_D$, which is $\frak{so}(2,D)$, and which acts on the Dirac cone embedded in a $D+2$ dimensional space \cite{Brust:2016gjy}.  Upon restriction to the $\frak{so}(1,D)$ isometry algebra of dS$_D$, this conformal representation splits into the representations in \eqref{conformreop}.

\section{Noether Currents\label{noethersec}}

In this section we give the Noether currents for the shift symmetric fields \eqref{shifssmgeee} under the shift symmetries \eqref{shifttransformation}.

The Lagrangian for a free massive spin-$s$ field on dS$_D$ is quite complicated, involving not only the field itself but also auxiliary fields that vanish on shell \cite{Hallowell:2005np}.  However, we are only interested in the Noether current arising from the shift symmetry, and its on shell structure is constrained enough that it can be pinned down using only general properties of the Lagrangian.  The canonically normalized Lagrangian depends only on the fields and their first derivatives and can be written in the form
\be {\cal L}=-\frac{1}{2}\nabla_\mu {\phi}_{\mu_1\cdots\mu_s}\nabla^\mu{\phi}^{\mu_1\cdots\mu_s}+({\rm on\ shell\ trivial\ terms})+({\rm zero\ derivative\ terms})\,.\label{massiveslagnele}\ee
 
The Noether current is linear in the field $\phi_{\mu_1\cdots\mu_s}$ (since the symmetry transformation does not involve the field and the action is quadratic in the field), linear in the variation $\delta\phi_{\mu_1\cdots\mu_s}$, and contains a single derivative (since the action is second order in derivatives).   The single derivative cannot contract with the field (due to the on shell divergencelessness of the field), and so it must carry the free index of the current.  These restrictions leave only two possible terms, with the relative coefficient fixed by ensuring that $\nabla_\mu \mathcal{J}^\mu =0$ on shell, given that both $\phi$ and $\delta \phi$ satisfy the on shell equations \eqref{eqnofmotion}.   This gives the following as the Noether current:
\begin{equation}
    \mathcal{J}^\mu = {\phi}_{\mu_1\cdots\mu_s} \nabla^\mu \delta {\phi}^{\mu_1\cdots\mu_s}  -   \nabla^\mu {\phi}_{\mu_1\cdots\mu_s} \,\delta {\phi}^{\mu_1\cdots\mu_s}\,.\label{spinsnotrhe}
\end{equation}
 The overall normalization is fixed by the Lagrangian: from the definition of the Noether current \eqref{exprnere}, the only term in \eqref{massiveslagnele} that contributes to the on shell current is the first term shown, and this gives \eqref{spinsnotrhe}.
 
This Noether current is given in terms of the symmetry transformation \eqref{shifttransformation}, which is parametrized by the tensor \be
S_{A_1\cdots A_{s+k},\, B_1\cdots B_s} \in  ~\raisebox{1.15ex}{\gyoung(_5{s+k},_3{s})}\,.\label{nothebnnree}
\ee
Note that the transformation \eqref{shifttransformation} is itself a shift by the mode \eqref{modesdefe} corresponding to this tensor,
\be  \delta_k{\phi}^{(s,k)}_{\mu_1\cdots\mu_s} ={\zeta}^{(s,s+k)}_{\mu_1\cdots\mu_s} \,.\ee

\section{Correspondence\label{conjecturesection}}

We can see now from the classification of CKTs in section \ref{CKTsection} that for a given $s$ and $k$ there will be many more currents that can be constructed from the CKTs than there are Noether currents for the shift symmetry.  A spin $s$ level $k$ field will have the rank $s+k+1$ field strength \eqref{fieldstrnree}, which contracts with a rank $r=s+k$ CKT to form the currents \eqref{fjojccree}.  The CKTs split into branches $l=0,1,\ldots, s+k$, parametrized by embedding space tensors as in \eqref{conformreop} with $r=s+k$.   The shift symmetries are only parametrized by one embedding space tensor, \eqref{nothebnnree}, and in the list of the CKT modes the only one that corresponds to this is the one at $l=s$.  We are thus led to conjecture that all the remaining currents are trivial, and the only non-trivial one is the one at $l=s$, which is equivalent to the Noether current \eqref{spinsnotrhe}.  We can illustrate this as follows:
\be \overset{l=s+k}{ ~\raisebox{1.15ex}{\gyoung(_5{s+k},_5{s+k})}}\,, \ \cdots \  , \  \overset{l=s+1}{ ~\raisebox{1.15ex}{\gyoung(_5{s+k},_4{s+1})} }\,, \  \boxed{\overset{l=s}{ ~\raisebox{1.15ex}{\gyoung(_5{s+k},_3{s})}} }\, ,  \  \overset{l=s-1}{ ~\raisebox{1.15ex}{\gyoung(_5{s+k},_3{s-1})}} \, , \ \cdots \ \,  , \   \overset{l=0}{ ~\raisebox{1.15ex}{\gyoung(_5{s+k})} } \,. \ee
These represent all the currents coming from CKTs: the boxed currents at $l=s$ are equivalent to the Noether currents, and the rest are trivial.

\begin{figure}
    \centering
    \begin{tikzpicture}[
    hgrid/.style={orange, very thick},
    vgrid/.style={green!50!black, very thick},
    dgrid/.style={blue, very thick},
    hlabel/.style={orange, font=\sffamily\bfseries},
    vlabel/.style={green!50!black, font=\sffamily\bfseries},
    dlabel/.style={blue, font=\sffamily\bfseries}
]

    \def\maxrow{5}

    \foreach \r in {0,...,\maxrow} {
     
        \draw[hgrid] (-1.3, -\r) -- (\r, -\r);

        \node[hlabel, anchor=east] at (-1.5, -\r) {$r=\r$};
    }

    \foreach \c in {0,...,\maxrow} {
        
        \draw[vgrid] (\c, -\c + 1.3) -- (\c, -\maxrow - 0.8);

        \node[vlabel, anchor=south] at (\c, -\c + 1.5) {$s=l=\c$};
    }
    
    \draw[dgrid] (-1.25, 1.25) -- (\maxrow, -\maxrow);
    \node[dlabel, anchor=south east] at (-1.25, 1.25) {$k=0$};

    \foreach \k in {1,...,\maxrow} {
        
        \draw[dgrid] (-1.25, -\k + 1.25) -- (\maxrow - \k + 0.8, -\maxrow - 0.8);

        \node[dlabel, anchor=south east] at (-1.25, -\k + 1.25) {$k=\k$};
    }

      \foreach \r in {0,...,\maxrow} {
        \foreach \c in {0,...,\r} {
            \fill[red] (\c, -\r) circle (1.5pt);
        }
    }

    \node[black, font=\Huge] at (\maxrow/2, -\maxrow - 1.5) {$\vdots$};

\end{tikzpicture}
    \caption{\small Matching of non-trivial currents to shift symmetries.  Each red dot represents a different Killing tensor mode.  The various lines indicate $r,l$ values of the Killing tensor mode, as well as the $s,k$ values of the shift symmetric field for which it occurs non-trivially. }
    \label{crossingfig}
\end{figure}
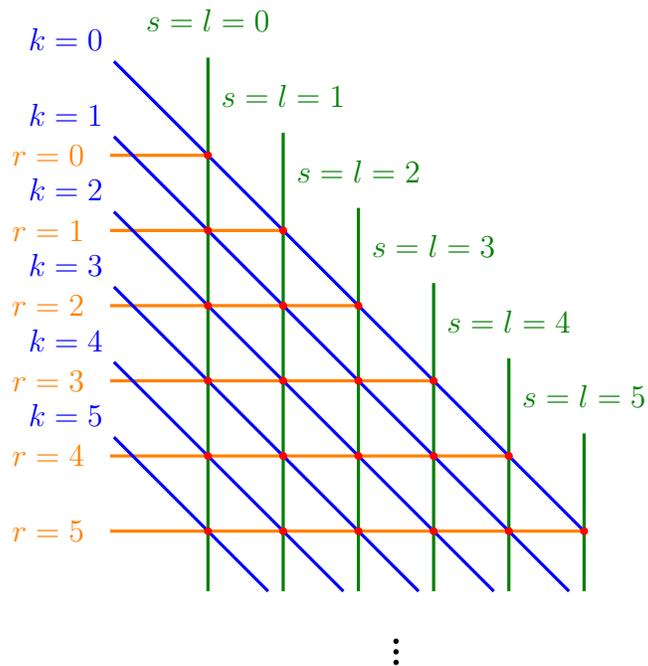

As we move through the spins and the values of $k$, each possible CKT tensor gets used exactly once as a non-trivial current for a shift symmetry.  This matching is illustrated in figure \ref{crossingfig}.

\section{Examples\label{examplessec}}

Though we do not have a full proof of the correspondence just described, we can check it in several of the first non-trivial cases\footnote{An analogous correspondence should also hold for the more general mixed symmetry \cite{Hinterbichler:2022vcc} and fermionic \cite{Bonifacio:2023prb} shift symmetric fields.  These presumably involve generalized Killing-Yano tensors (see \cite{Hinterbichler:2022agn} for further references), and their spinor versions.}

The $s=0$, $k=0$ case is the trivial case, the only case where there are no trivial currents: the shift symmetry is a simple shift $\delta \phi=S$ with $S$ constant, and the Noether current is ${\cal J}^\mu=-S\nabla^\mu\phi$, which is proportional to the field strength $F_\mu=\nabla_\mu\phi$ multiplied by the trivial $r=l=0$ conformal Killing tensor $S$.  The $s=0$, $k=1$ case is the case described in the introduction.

For the higher cases, the improvement terms needed to see the matching proliferate, so in what follows we introduce some notation.
A CKT current for the spin $s$ level $k$ field $\tensor{\phi}{^{^{(s,k)}}}{_{\mu_{1}\cdots\mu_s}}$, constructed from an $l$-th mode CKT using \eqref{fjojccree}, \eqref{fieldstrnree}, will be denoted
\bea
     J^{(s,k)(l)}_\mu   &=&\nabla_{(\mu}\nabla_{\mu_1}\cdots\nabla_{\mu_{k}}\tensor{\phi}{^{^{(s,k)}}}{_{\mu_{k+1}\cdots\mu_{s+k})_T}}\nabla^{(\mu_1}\cdots\nabla^{\mu_{s+k-l}}\tensor{\zeta}{^{^{(l,s+k)}}}{^{\mu_{s+k-l+1}\cdots\mu_{s+k})_T}}\,, \label{Jskdefnne}
\eea
where we have written the CKT in terms of the mode using \eqref{symconvee} and the requirement $r=s+k$.

There is a current that is identically equal to zero due to the conformal Killing equations \eqref{conforkee2},
\bea
    {K}^{(s,k)(l)}_\mu&=&\nabla^{(\mu_1}\cdots\nabla^{\mu_{k}}\tensor{\phi}{^{^{(s,k)}}}{^{\mu_{k+1}\cdots\mu_{s+k})_T}}\nabla_{(\mu} \tensor{\xi}{^{^{(l,s+k)}}}{_{\mu_1\cdots\mu_{s+k})_T}} \nn\\
    &=&\nabla^{(\mu_1}\cdots\nabla^{\mu_{k}}\tensor{\phi}{^{^{(s,k)}}}{^{\mu_{k+1}\cdots\mu_{s+k})_T}}\nabla_{(\mu}\nabla_{\mu_1}\cdots\nabla_{\mu_{s+k-l}} \tensor{\zeta}{^{^{(l,s+k)}}}{_{\mu_{s+k-l+1}\cdots\mu_{s+k})_T}}\,.
\eea
It will be useful to add this to aid in seeing some of the identities that follow; it is of course not strictly necessary since it is vanishing, but the identities will be easier to see when it is included.

We then look to add improvement terms to the combination
\begin{equation}
    {J}^{(s,k)(l)}_\mu-{K}^{(s,k)(l)}_\mu\,,
\end{equation}
to either make it vanish or to make it proportional to the Noether current \eqref{spinsnotrhe}. We can parametrize the possible improvement terms as follows.  First of all, the improvement terms must use rank 2 anti-symmetric tensors, and they must contain one power of the field and one power of the mode tensor.  On top of this, they can have more powers of the derivatives.  The derivatives cannot contract with the fields or mode tensor because both are transverse.  Furthermore, derivatives acting on one can only contract with derivatives acting on the other, since both satisfy Klein-Gordon equations.  There is also an upper limit on the number of derivatives, set by the number of derivatives in ${J}^{(s,k)(l)}_\mu$ in \eqref{Jskdefnne}.  Given these restrictions, there will be a finite basis of possible improvement terms in each case.
A basis of improvement terms will be denoted
\begin{equation}
   \mathcal{I}_{(a,a')\,\mu\nu}^{(s,k)(l)}\, .
\end{equation}
The basis is indexed by the labels $a$ and $a'$: $a$ is a label that gives the number of pairs of contracted indices in the term, and $a'$ is an additional label to distinguish distinct terms with the same number of contracted indices.  

For example, when $s=l=0$ both the field and the CKT mode are scalars and there is only one possible contraction at each derivative order.  We have $a\leq k-1$, there is no need for an $a'$ index, and we can choose a basis as follows: 
\begin{align}
    &\mathcal{I}_{(0)\,\mu\nu}^{(0,k)(0)}=\nabla_{[\mu}{\phi}^{(0,k)}\nabla_{\nu]}{\zeta}^{(0,k)} \,,\nn\\
    &\mathcal{I}_{(1)\,\mu\nu}^{(0,k)(0)} = \nabla^{\mu_1}\nabla_{[\mu|}{\phi}^{(0,k)}\nabla_{\mu_1}\nabla_{|\nu]}{\zeta}^{(0,k)}\,,\nn\\
    &\vdots \nn\\
    &\mathcal{I}_{(k-1)\,\mu\nu}^{(0,k)(0)}= \nabla^{\mu_{1}}\cdots\nabla^{\mu_{k-1}}\nabla_{[\mu|}{\phi}^{(0,k)}\nabla_{\mu_{1}}\cdots\nabla_{\mu_{k-1}}\nabla_{|\nu]}{\zeta}^{(0,k)}\,.
\end{align}
When $s=0$ and $l>0$, the restriction on $a$ is $l-1 \leq a \leq k-1$ and we have two possible terms at each derivative order, which are distinguished by $a'=1,2$.  We can choose the basis as 
\begin{align}
    &\mathcal{I}_{(l-1,0){\mu\nu}}^{(0,k)(l)} = \nabla^{\mu_{1}}\cdots\nabla^{\mu_{l-1}}\tensor{\phi}{^{^{(0,k)}}}\nabla_{[\mu}\tensor{\zeta}{^{^{(l,k)}}}{_{\nu]\mu_1\cdots\mu_{l-1}}}\,,\nn\\
    &\mathcal{I}_{(l-1,1){\mu\nu}}^{(0,k)(l)} = \nabla^{\mu_{1}}\cdots\nabla^{\mu_{l-1}}\nabla_{[\mu}\tensor{\phi}{^{^{(0,k)}}}\tensor{\zeta}{^{^{(l,k)}}}{_{\nu]\mu_1\cdots\mu_{l-1}}}\,,\nn\\
    &\vdots\nn\\
    &\mathcal{I}_{(k-1,0){\mu\nu}}^{(0,k)(l)} = \nabla^{\mu_{1}}\cdots\nabla^{\mu_{k-1}}\tensor{\phi}{^{^{(0,k)}}}\nabla_{\mu_{1}}\cdots\nabla_{\mu_{k-l}}\nabla_{[\mu}\tensor{\zeta}{^{^{(l,k)}}}{_{\nu]\mu_{k-l+1}\cdots\mu_{k-1}}}\,,\nn\\
    &\mathcal{I}_{(k-1,1){\mu\nu}}^{(0,k)(l)} = \nabla^{\mu_{1}}\cdots\nabla^{\mu_{k-1}}\nabla_{[\mu|}\tensor{\phi}{^{^{(0,k)}}}\nabla_{\mu_{1}}\cdots\nabla_{\mu_{k-l}}\tensor{\zeta}{^{^{(l,k)}}}{_{|\nu]\mu_{k-l+1}\cdots\mu_{k-1}}}\,.
\end{align}
When $l=0$ and $s>0$, the terms are similar, with $\phi$ and $\zeta$ and their corresponding labels swapped and the restriction $s-1 \leq a \leq s+k-1$,
\begin{align}
    &\mathcal{I}_{(s-1,0)\mu\nu}^{(s,k)(0)} = \nabla^{\mu_{1}}\cdots\nabla^{\mu_{s-1}}\tensor{\zeta}{^{^{(0,k)}}}\nabla_{[\mu}\tensor{\phi}{^{^{(s,k)}}}{_{\nu]\mu_1\cdots\mu_{s-1}}}\,,\nn\\
    &\mathcal{I}_{(s-1,1)\mu\nu}^{(s,k)(0)} = \nabla^{\mu_{1}}\cdots\nabla^{\mu_{s-1}}\nabla_{[\mu}\tensor{\zeta}{^{^{(0,k)}}}\tensor{\phi}{^{^{(s,k)}}}{_{\nu]\mu_1\cdots\mu_{s-1}}}\,,\nn\\
    &\vdots\nn\\
    &\mathcal{I}_{(s+k-1,0)\mu\nu}^{(s,k)(0)}  = \nabla^{\mu_{1}}\cdots\nabla^{\mu_{s+k-1}}\tensor{\zeta}{^{^{(0,k)}}}\nabla_{\mu_{1}}\cdots\nabla_{\mu_k}\nabla_{[\mu}\tensor{\phi}{^{^{(s,k)}}}{_{\nu]\mu_{k+1}\cdots\mu_{s+k-1}}}\,,\nn\\
    &\mathcal{I}_{(s+k-1,1)\mu\nu}^{(s,k)(0)}  = \nabla^{\mu_{1}}\cdots\nabla^{\mu_{s+k-1}}\nabla_{[\mu|}\tensor{\zeta}{^{^{(0,k)}}}\nabla_{\mu_{1}}\cdots\nabla_{\mu_{k}}\tensor{\phi}{^{^{(s,k)}}}{_{|\nu]\mu_{k+1}\cdots\mu_{s+k-1}}}\,.
\end{align}

For each field $(s,k)$ there is only one Noether Current $\mathcal{J}^{(s,k)}_\mu$ but $s+k+1$ classes of CKT currents ${J}^{(s,k)(l)}_\mu$, distinguished by the different choices of $l=0,1,\cdots,s+k$. 
Our conjecture from section \ref{conjecturesection} says that for $l=s$ the CKT currents will be proportional to the Noether current up to improvement terms, and the CKT currents for $l\not=s$ will be pure improvement terms.
Altogether, we should thus have
\begin{equation}
    {J}^{(s,k)(l)}_\mu-{K}^{(s,k)(l)}_\mu = 
    \begin{cases}
      \sum_{a,a'}  c_{a,a'} \nabla^\nu \mathcal{I}_{(a,a')\mu\nu}^{(s,k)(l)} \, , & \text{if $l\neq s$}  \, , \\
      \sum_{a,a'}  c_{a,a'}  \nabla^\nu \mathcal{I}_{(a,a')\mu\nu}^{(s,k)(l)}+ c{\mathcal{J}}^{(s,k)}_\mu \, , & \text{if $l=s$} \, ,
    \end{cases}       \label{cosntntee}
\end{equation}
where  $c_{a,a'}$ and $c$ are constants. 

We have checked explicitly that this works for $s=\{0,1,2\}$ with $r=\{1,2,3\}$. Results for the constants in \eqref{cosntntee} are shown in tables \ref{tab:numerical}, \ref{tab:numerical2} and \ref{tab:numerical3}.\footnote{Note that since ${J}^{(s,k)(l)}_\mu$ has the same form as ${K}^{(l,r-l)(s)}_\mu$, and ${K}^{(s,k)(l)}_\mu$ has the same form as ${J}^{(l,r-l)(s)}_\mu$, the cases $(s,k)(l)$ and $(l,r-l)(s)$ have improvement terms of the same forms with $\phi$ and $\zeta$ swapped, with coefficients that are the same up to an overall sign once the order of the anti-symmetrization $[\mu, \nu]$ is accounted for.} 

\begin{longtable}{|p{30pt}|p{30pt}|p{230pt}|p{150pt}|}
    \hline
      k & l & Improvement Terms & Constants\\\hline
      \multirow{2}{*}{1} 
      & 0 & 
      $\mathcal{I}_{(0)\mu\nu}^{(0,1)(0)}=\nabla_{[\mu}\tensor{\phi}{^{^{(0,1)}}}\nabla_{\nu]}\tensor{\zeta}{^{^{(0,1)}}}$ 
      &
      $c=(1-D)H^2$
      \newline
      $c_{0}=-2$
      \\\cline{2-4}
      & 1 & 
      $\mathcal{I}_{(0,0)\mu\nu}^{(0,1)(1)}=\tensor{\phi}{^{^{(0,1)}}}\nabla_{[\mu}\tensor{\zeta}{^{^{(1,1)}}}{_{\nu]}}
      \newline 
      \mathcal{I}_{(0,1)\mu\nu}^{(0,1)(1)}=\nabla_{[\mu}\tensor{\phi}{^{^{(0,1)}}}\tensor{\zeta}{^{^{(1,1)}}}{_{\nu]}}$ 
      &
      $c_{0,0}=1
      \newline
      c_{0,1}=-2$
      \\\hline
      \multirow{3}{*}{2} 
      & 0 &
      $\mathcal{I}_{(0)\mu\nu}^{(0,2)(0)}=\nabla_{[\mu}\tensor{\phi}{^{^{(0,2)}}}\nabla_{\nu]}\tensor{\zeta}{^{^{(0,2)}}} 
      \newline 
      \mathcal{I}_{(1)\mu\nu}^{(0,2)(0)}=\nabla^{\mu_1}\nabla_{[\mu|}\tensor{\phi}{^{^{(0,2)}}}\nabla_{\mu_1}\nabla_{|\nu]}\tensor{\zeta}{^{^{(0,2)}}}$
      &
      $c=2(1-D^2)H^4$
      \newline
      $c_{0}=-2DH^2 
      \newline
      c_{1}=-2$ 
      \\\cline{2-4}
      & 1 & 
      $\mathcal{I}_{(0,0)\mu\nu}^{(0,2)(1)} = \tensor{\phi}{^{^{(0,2)}}}\nabla_{[\mu}\tensor{\zeta}{^{^{(1,2)}}}{_{\nu]}} 
      \newline
      \mathcal{I}_{(0,1)\mu\nu}^{(0,2)(1)} = \nabla_{[\mu}\tensor{\phi}{^{^{(0,2)}}}\tensor{\zeta}{^{^{(1,2)}}}{_{\nu]}} 
      \newline
      \mathcal{I}_{(1,0)\mu\nu}^{(0,2)(1)} = \nabla^{\mu_1}\tensor{\phi}{^{^{(0,2)}}}\nabla_{\mu_1}\nabla_{[\mu}\tensor{\zeta}{^{^{(1,2)}}}{_{\nu]}}
      \newline 
      \mathcal{I}_{(1,1)\mu\nu}^{(0,2)(1)} = \nabla^{\mu_1}\nabla_{[\mu|}\tensor{\phi}{^{^{(0,2)}}}\nabla_{\mu_1}\tensor{\zeta}{^{^{(1,2)}}}{_{|\nu]}}$ 
      &
      $c_{0,0}=-\frac{4}{3}(1+D)H^2 
      \newline
      c_{0,1}=2DH^2 
      \newline
      c_{1,0}=\frac{4}{3} 
      \newline
      c_{1,1}=-2 $ 
      \\\cline{2-4}
      & 2 &
      $\mathcal{I}_{(1,0)\mu\nu}^{(0,2)(2)} = \nabla^{\mu_1}\tensor{\phi}{^{^{(0,2)}}}\nabla_{[\mu}\tensor{\zeta}{^{^{(2,2)}}}{_{\nu]\mu_1}}
      \newline 
      \mathcal{I}_{(1,1)\mu\nu}^{(0,2)(2)} = \nabla^{\mu_1}\nabla_{[\mu}\tensor{\phi}{^{^{(0,2)}}}\tensor{\zeta}{^{^{(2,2)}}}{_{\nu]\mu_1}}$ 
      &
      $c_{1,0}=\frac{2}{3}
      \newline
      c_{1,1}=-2$ 
      \\\hline
      \multirow{4}{*}{3}
      & 0 &
      $\mathcal{I}_{(0)\mu\nu}^{(0,3)(0)} =\nabla_{[\mu}\tensor{\phi}{^{^{(0,3)}}}\nabla_{\nu]}\tensor{\zeta}{^{^{(0,3)}}}
      \newline
      \mathcal{I}_{(1)\mu\nu}^{(0,3)(0)} =\nabla^{\mu_1}\nabla_{[\mu|}\tensor{\phi}{^{^{(0,3)}}}\nabla_{\mu_1}\nabla_{|\nu]}\tensor{\zeta}{^{^{(0,3)}}}
      \newline
      \mathcal{I}_{(2)\mu\nu}^{(0,3)(0)} =\nabla^{\mu_1}\nabla^{\mu_2}\nabla_{[\mu|}\tensor{\phi}{^{^{(0,3)}}}\nabla_{\mu_1}\nabla_{\mu_2}\nabla_{|\nu]}\tensor{\zeta}{^{^{(0,3)}}}$
      &
      $c=6  (3 + D - 3 D^2 - D^3) H^6 $
      \newline
      $c_{0}=-4(-9+D^2)H^4
      \newline
      c_{1}=-2(1+D)H^2
      \newline 
      c_{2}=-2$
      \\\cline{2-4}
      & 1 &
      $\mathcal{I}_{(0,0)\mu\nu}^{(0,3)(1)} = \tensor{\phi}{^{^{(0,3)}}}\nabla_{[\mu}\tensor{\zeta}{^{^{(1,3)}}}{_{\nu]}}
      \newline
      \mathcal{I}_{(0,1)\mu\nu}^{(0,3)(1)} = \nabla_{[\mu}\tensor{\phi}{^{^{(0,3)}}}\tensor{\zeta}{^{^{(1,3)}}}{_{\nu]}}
      \newline
      \mathcal{I}_{(1,0)\mu\nu}^{(0,3)(1)} = \nabla^{\mu_1}\tensor{\phi}{^{^{(0,3)}}}\nabla_{\mu_1}\nabla_{[\mu}\tensor{\zeta}{^{^{(1,3)}}}{_{\nu]}}
      \newline
      \mathcal{I}_{(1,1)\mu\nu}^{(0,3)(1)} = \nabla^{\mu_1}\nabla_{[\mu|}\tensor{\phi}{^{^{(0,3)}}}\nabla_{\mu_1}\tensor{\zeta}{^{^{(1,3)}}}{_{|\nu]}}
      \newline
      \mathcal{I}_{(2,0)\mu\nu}^{(0,3)(1)} = \nabla^{\mu_1}\nabla^{\mu_2}\tensor{\phi}{^{^{(0,3)}}}\nabla_{\mu_1}\nabla_{\mu_2}\nabla_{[\mu}\tensor{\zeta}{^{^{(1,3)}}}{_{\nu]}}
      \newline
      \mathcal{I}_{(2,1)\mu\nu}^{(0,3)(1)} = \nabla^{\mu_1}\nabla^{\mu_2}\nabla_{[\mu|}\tensor{\phi}{^{^{(0,3)}}}\nabla_{\mu_1}\nabla_{\mu_2}\tensor{\zeta}{^{^{(1,3)}}}{_{|\nu]}}$ 
      &
      $c_{0,0}=\frac{3}{2}(-16-3D+2D^2)H^4
      \newline
      c_{0,1}=2(11+5D-2D^2)H^4
      \newline
      c_{1,0}=-\frac{1}{2}(11+6D)H^2
      \newline
      c_{1,1}=4(1+D)H^2
      \newline
      c_{2,0}=\frac{3}{2}
      \newline
      c_{2,1}=-2$
      \\\cline{2-4}
      & 2 & 
      $\mathcal{I}_{(1,0)\mu\nu}^{(0,3)(2)} = \nabla^{\mu_1}\tensor{\phi}{^{^{(0,3)}}}\nabla_{[\mu}\tensor{\zeta}{^{^{(2,3)}}}{_{\nu]\mu_1}}
      \newline
      \mathcal{I}_{(1,1)\mu\nu}^{(0,3)(2)} = \nabla^{\mu_1}\nabla_{[\mu}\tensor{\phi}{^{^{(0,3)}}}\tensor{\zeta}{^{^{(2,3)}}}{_{\nu]\mu_1}}
      \newline
      \mathcal{I}_{(2,0)\mu\nu}^{(0,3)(2)} = \nabla^{\mu_1}\nabla^{\mu_2}\tensor{\phi}{^{^{(0,3)}}}\nabla_{\mu_1}\nabla_{[\mu}\tensor{\zeta}{^{^{(2,3)}}}{_{\nu]\mu_2}}
      \newline
      \mathcal{I}_{(2,1)\mu\nu}^{(0,3)(2)} = \nabla^{\mu_1}\nabla^{\mu_2}\nabla_{[\mu|}\tensor{\phi}{^{^{(0,3)}}}\nabla_{\mu_1}\tensor{\zeta}{^{^{(2,3)}}}{_{|\nu]\mu_2}}$ 
      &
      $c_{1,0}=-(3+D)H^2
      \newline
      c_{1,1}=2(1+D)H^2
      \newline
      c_{2,0}=1
      \newline
      c_{2,1}=-2$
      \\\cline{2-4}
      & 3 &
      $\mathcal{I}_{(2,0)\mu\nu}^{(0,3)(3)} = \nabla^{\mu_1}\nabla^{\mu_2}\tensor{\phi}{^{^{(0,3)}}}\nabla_{[\mu}\tensor{\zeta}{^{^{(3,3)}}}{_{\nu]\mu_1\mu_2}}
      \newline 
      \mathcal{I}_{(2,1)\mu\nu}^{(0,3)(3)} = \nabla^{\mu_1}\nabla^{\mu_2}\nabla_{[\mu}\tensor{\phi}{^{^{(0,3)}}}\tensor{\zeta}{^{^{(3,3)}}}{_{\nu]\mu_1\mu_2}}$ 
      &
      $c_{2,0}=\frac{1}{2}
      \newline
      c_{2,1}=-2$
      \\\hline
      \caption{Improvement terms for the shift symmetric scalars with $k=1,2,3$.}
        \label{tab:numerical}
\end{longtable}

\begin{longtable}{|p{30pt}|p{30pt}|p{230pt}|p{150pt}|}
    \hline
      k & l & Improvement Terms & Constants\\\hline
      \multirow{2}{*}{0} 
      & 0 & 
      $\mathcal{I}_{(0,0)\mu\nu}^{(1,0)(0)}=\nabla_{[\mu}\tensor{\phi}{^{^{(1,0)}}}{_{\nu]}}\tensor{\zeta}{^{^{(0,1)}}}$ 
      \newline
      $\mathcal{I}_{(0,1)\mu\nu}^{(1,0)(0)}=\tensor{\phi}{^{^{(1,0)}}}{_{[\mu}}\nabla_{\nu]}\tensor{\zeta}{^{^{(0,1)}}}$
      &
      $c_{0,0} = -1$
      \newline
      $c_{0,1} = -2$
      \\\cline{2-4}
      & 1 &
      $\mathcal{I}_{(0)\mu\nu}^{(1,0)(1)}=\tensor{\phi}{^{^{(1,0)}}}{_{[\mu}}\tensor{\zeta}{^{^{(1,1)}}}{_{\nu]}}$ 
      &
      $c = -\frac{1}{2}$
      \newline
      $c_{0} = -1$
      \\\hline
      \multirow{3}{*}{1}
      & 0 &
      $\mathcal{I}_{(0,0)\mu\nu}^{(1,1)(0)}=\nabla_{[\mu}\tensor{\phi}{^{^{(1,1)}}}{_{\nu]}}\tensor{\zeta}{^{^{(0,2)}}}$ 
      \newline
      $\mathcal{I}_{(0,1)\mu\nu}^{(1,1)(0)}=\tensor{\phi}{^{^{(1,1)}}}{_{[\mu}}\nabla_{\nu]}\tensor{\zeta}{^{^{(0,2)}}}$
      \newline
      $\mathcal{I}_{(1,0)\mu\nu}^{(1,1)(0)}=\nabla^{\mu_1}\nabla_{[\mu}\tensor{\phi}{^{^{(1,1)}}}{_{\nu]}}\nabla_{\mu_1}\tensor{\zeta}{^{^{(0,2)}}}$ 
      \newline
      $\mathcal{I}_{(1,1)\mu\nu}^{(1,1)(0)}=\nabla^{\mu_1}\tensor{\phi}{^{^{(1,1)}}}{_{[\mu|}}\nabla_{\mu_1}\nabla_{|\nu]}\tensor{\zeta}{^{^{(0,2)}}}$
      &
      $c_{0,0} = \frac{4}{3}(1+D)H^2$
      \newline
      $c_{0,1} = 2DH^2$
      \newline
      $c_{1,0} = -\frac{4}{3}$
      \newline
      $c_{1,1} = -2$
      \\\cline{2-4}
      & 1 &
      $\mathcal{I}_{(0)\mu\nu}^{(1,1)(1)}=\tensor{\phi}{^{^{(1,1)}}}{_{[\mu}}\tensor{\zeta}{^{^{(1,2)}}}{_{\nu]}}$
      \newline
      $\mathcal{I}_{(1,0)\mu\nu}^{(1,1)(1)}=\nabla_{[\mu}\tensor{\phi}{^{^{(1,1)}}}{^{\mu_1}}\nabla_{\nu]}\tensor{\zeta}{^{^{(1,2)}}}{_{\mu_1}}$
      \newline
      $\mathcal{I}_{(1,1)\mu\nu}^{(1,1)(1)}=\nabla_{[\mu|}\tensor{\phi}{^{^{(1,1)}}}{^{\mu_1}}\nabla_{\mu_1}\tensor{\zeta}{^{^{(1,2)}}}{_{|\nu]}}$
      \newline
      $\mathcal{I}_{(1,2)\mu\nu}^{(1,1)(1)}=\nabla^{\mu_1}\tensor{\phi}{^{^{(1,1)}}}{_{[\mu}}\nabla_{\nu]}\tensor{\zeta}{^{^{(1,2)}}}{_{\mu_1}}$
      &
      $c = -\frac{1}{3}(1+D)H^2$
      \newline
      $c_{0} = \frac{2}{3}(1-2D)H^2$
      \newline
      $c_{1,0} = -\frac{2}{3}$
      \newline
      $c_{1,1} = -\frac{2}{3}$
      \newline
      $c_{1,2} = -\frac{2}{3}$
      \\\cline{2-4}
      & 2 &
      $\mathcal{I}_{(1,0)\mu\nu}^{(1,1)(2)}=\nabla_{[\mu}\tensor{\phi}{^{^{(1,1)}}}{^{\mu_1}}\tensor{\zeta}{^{^{(2,2)}}}{_{\nu]\mu_1}}$
      \newline
      $\mathcal{I}_{(1,1)\mu\nu}^{(1,1)(2)}=\tensor{\phi}{^{^{(1,1)}}}{^{\mu_1}}\nabla_{[\mu}\tensor{\zeta}{^{^{(2,2)}}}{_{\nu]\mu_1}}$
      \newline
      $\mathcal{I}_{(1,2)\mu\nu}^{(1,1)(2)}=\nabla^{\mu_1}\tensor{\phi}{^{^{(1,1)}}}{_{[\mu}}\tensor{\zeta}{^{^{(2,2)}}}{_{\nu]\mu_1}}$
      &
      $c_{1,0} = -\frac{4}{3}$
      \newline
      $c_{1,1} = \frac{2}{3}$
      \newline
      $c_{1,2} = -\frac{2}{3}$
      \\\hline
      \multirow{4}{*}{2}
      & 0 &
      $\mathcal{I}_{(0,0)\mu\nu}^{(1,2)(0)}=\nabla_{[\mu}\tensor{\phi}{^{^{(1,2)}}}{_{\nu]}}\tensor{\zeta}{^{^{(0,3)}}}$ 
      \newline
      $\mathcal{I}_{(0,1)\mu\nu}^{(1,2)(0)}=\tensor{\phi}{^{^{(1,2)}}}{_{[\mu}}\nabla_{\nu]}\tensor{\zeta}{^{^{(0,3)}}}$
      \newline
      $\mathcal{I}_{(1,0)\mu\nu}^{(1,2)(0)}=\nabla^{\mu_1}\nabla_{[\mu}\tensor{\phi}{^{^{(1,2)}}}{_{\nu]}}\nabla_{\mu_1}\tensor{\zeta}{^{^{(0,3)}}}$ 
      \newline
      $\mathcal{I}_{(1,1)\mu\nu}^{(1,2)(0)}=\nabla^{\mu_1}\tensor{\phi}{^{^{(1,2)}}}{_{[\mu|}}\nabla_{\mu_1}\nabla_{|\nu]}\tensor{\zeta}{^{^{(0,3)}}}$
      \newline
      $\mathcal{I}_{(2,0)\mu\nu}^{(1,2)(0)}=\nabla^{\mu_1}\nabla^{\mu_2}\nabla_{[\mu}\tensor{\phi}{^{^{(1,2)}}}{_{\nu]}}\nabla_{\mu_1}\nabla_{\mu_2}\tensor{\zeta}{^{^{(0,3)}}}$ 
      \newline
      $\mathcal{I}_{(2,1)\mu\nu}^{(1,2)(0)}=\nabla^{\mu_1}\nabla^{\mu_2}\tensor{\phi}{^{^{(1,2)}}}{_{[\mu|}}\nabla_{\mu_1}\nabla_{\mu_2}\nabla_{|\nu]}\tensor{\zeta}{^{^{(0,3)}}}$
      &
      $c_{0,0} = \frac{3}{2}(16+3D-2D^2)H^4$
      \newline
      $c_{0,1} = 2(11+5D-2D^2)H^4$
      \newline
      $c_{1,0} = \frac{1}{2}(11+6D)H^2$
      \newline
      $c_{1,1} = 4(1+D)H^2$
      \newline
      $c_{2,0} = -\frac{3}{2}$
      \newline
      $c_{2,1} = -2$
       \\\cline{2-4}
      & 1 &
      $\mathcal{I}_{(0)\mu\nu}^{(1,2)(1)}=\tensor{\phi}{^{^{(1,2)}}}{_{[\mu}}\tensor{\zeta}{^{^{(1,3)}}}{_{\nu]}}$
      \newline
      $\mathcal{I}_{(1,0)\mu\nu}^{(1,2)(1)}=\nabla_{[\mu}\tensor{\phi}{^{^{(1,2)}}}{^{\mu_1}}\nabla_{\nu]}\tensor{\zeta}{^{^{(1,3)}}}{_{\mu_1}}$
      \newline
      $\mathcal{I}_{(1,1)\mu\nu}^{(1,2)(1)}=\nabla_{[\mu|}\tensor{\phi}{^{^{(1,2)}}}{^{\mu_1}}\nabla_{\mu_1}\tensor{\zeta}{^{^{(1,3)}}}{_{|\nu]}}$
      \newline
      $\mathcal{I}_{(1,2)\mu\nu}^{(1,2)(1)}=\nabla^{\mu_1}\tensor{\phi}{^{^{(1,2)}}}{_{[\mu}}\nabla_{\nu]}\tensor{\zeta}{^{^{(1,3)}}}{_{\mu_1}}$
      \newline
      $\mathcal{I}_{(1,3)\mu\nu}^{(1,2)(1)}=\nabla^{\mu_1}\tensor{\phi}{^{^{(1,2)}}}{_{[\mu|}}\nabla_{\mu_1}\tensor{\zeta}{^{^{(1,3)}}}{_{|\nu]}}$
      \newline
      $\mathcal{I}_{(2,0)\mu\nu}^{(1,2)(1)}=\nabla^{\mu_1}\nabla_{[\mu|}\tensor{\phi}{^{^{(1,2)}}}{^{\mu_2}}\nabla_{\mu_1}\nabla_{|\nu]}\tensor{\zeta}{^{^{(1,3)}}}{_{\mu_2}}$
      \newline
      $\mathcal{I}_{(2,1)\mu\nu}^{(1,2)(1)}=\nabla^{\mu_1}\nabla_{[\mu|}\tensor{\phi}{^{^{(1,2)}}}{^{\mu_2}}\nabla_{\mu_1}\nabla_{\mu_2}\tensor{\zeta}{^{^{(1,3)}}}{_{|\nu]}}$
      \newline
      $\mathcal{I}_{(2,2)\mu\nu}^{(1,2)(1)}=\nabla^{\mu_1}\nabla^{\mu_2}\tensor{\phi}{^{^{(1,2)}}}{_{[\mu|}}\nabla_{\mu_1}\nabla_{|\nu]}\tensor{\zeta}{^{^{(1,3)}}}{_{\mu_2}}$
      \newline
      $\mathcal{I}_{(2,3)\mu\nu}^{(1,2)(1)}=\nabla^{\mu_1}\nabla^{\mu_2}\tensor{\phi}{^{^{(1,2)}}}{_{[\mu|}}\nabla_{\mu_1}\nabla_{\mu_2}\tensor{\zeta}{^{^{(1,3)}}}{_{|\nu]}}$
      &
      $c = -\frac{1}{2}(3+4D+D^2)H^4$
      \newline
      $c_{0} = (-1+4D+7D^2)H^4$
      \newline
      $c_{1,0} = -\frac{5}{2}H^2$
      \newline
      $c_{1,1} = (2+D)H^2$
      \newline
      $c_{1,2} = (2+D)H^2$
      \newline
      $c_{1,3} = -\frac{1}{2}(7+8D)H^2$
      \newline
      $c_{2,0} = -\frac{1}{2}$
      \newline
      $c_{2,1} = -1$
      \newline
      $c_{2,2} = -1$
      \newline
      $c_{2,3} = \frac{1}{2}$
      \\\cline{2-4}
      & 2 &
      $\mathcal{I}_{(1,0)\mu\nu}^{(1,2)(2)}=\nabla_{[\mu}\tensor{\phi}{^{^{(1,2)}}}{^{\mu_1}}\tensor{\zeta}{^{^{(2,3)}}}{_{\nu]\mu_1}}$
      \newline
      $\mathcal{I}_{(1,1)\mu\nu}^{(1,2)(2)}=\tensor{\phi}{^{^{(1,2)}}}{^{\mu_1}}\nabla_{[\mu}\tensor{\zeta}{^{^{(2,3)}}}{_{\nu]\mu_1}}$
      \newline
      $\mathcal{I}_{(1,2)\mu\nu}^{(1,2)(2)}=\nabla^{\mu_1}\tensor{\phi}{^{^{(1,2)}}}{_{[\mu}}\tensor{\zeta}{^{^{(2,3)}}}{_{\nu]\mu_1}}$
      \newline
      $\mathcal{I}_{(2,0)\mu\nu}^{(1,2)(2)}=\nabla^{\mu_1}\nabla_{[\mu|}\tensor{\phi}{^{^{(1,2)}}}{^{\mu_2}}\nabla_{\mu_1}\tensor{\zeta}{^{^{(2,3)}}}{_{|\nu]\mu_2}}$
      \newline
      $\mathcal{I}_{(2,1)\mu\nu}^{(1,2)(2)}=\nabla^{\mu_1}\tensor{\phi}{^{^{(1,2)}}}{^{\mu_2}}\nabla_{\mu_1}\nabla_{[\mu}\tensor{\zeta}{^{^{(2,3)}}}{_{\nu]\mu_2}}$
      \newline
      $\mathcal{I}_{(2,2)\mu\nu}^{(1,2)(2)}=\nabla^{\mu_1}\nabla^{\mu_2}\tensor{\phi}{^{^{(1,2)}}}{_{[\mu|}}\nabla_{\mu_1}\tensor{\zeta}{^{^{(2,3)}}}{_{|\nu]\mu_2}}$
      \newline
      $\mathcal{I}_{(2,3)\mu\nu}^{(1,2)(2)}=\nabla^{\mu_1}\nabla_{[\mu|}\tensor{\phi}{^{^{(1,2)}}}{^{\mu_2}}\nabla_{\mu_2}\tensor{\zeta}{^{^{(2,3)}}}{_{|\nu]\mu_1}}$
      \newline
      $\mathcal{I}_{(2,4)\mu\nu}^{(1,2)(2)}=\nabla^{\mu_1}\tensor{\phi}{^{^{(1,2)}}}{^{\mu_2}}\nabla_{\mu_2}\nabla_{[\mu}\tensor{\zeta}{^{^{(2,3)}}}{_{\nu]\mu_1}}$
      &
      $c_{1,0} = \frac{1}{6}(7+9D)H^2$
      \newline
      $c_{1,1} = -\frac{1}{2}(3+2D)H^2$
      \newline
      $c_{1,2} = \frac{1}{6}(5+3D)H^2$
      \newline
      $c_{2,0} = -1$
      \newline
      $c_{2,1} = \frac{2}{3}$
      \newline
      $c_{2,2} = -\frac{1}{2}$
      \newline
      $c_{2,3} = -\frac{1}{2}$
      \newline
      $c_{2,4} = \frac{1}{3}$
      \\\cline{2-4}
      & 3 &
      $\mathcal{I}_{(2,0)\mu\nu}^{(1,2)(3)}=\nabla^{\mu_1}\nabla^{\mu_2}\tensor{\phi}{^{^{(1,2)}}}{_{[\mu}}\tensor{\zeta}{^{^{(3,3)}}}{_{\nu]\mu_1\mu_2}}$
      \newline
      $\mathcal{I}_{(2,1)\mu\nu}^{(1,2)(3)}=\nabla^{\mu_1}\nabla_{[\mu}\tensor{\phi}{^{^{(1,2)}}}{^{\mu_2}}\tensor{\zeta}{^{^{(3,3)}}}{_{\nu]\mu_1\mu_2}}$
      \newline
      $\mathcal{I}_{(2,2)\mu\nu}^{(1,2)(3)}=\nabla^{\mu_1}\tensor{\phi}{^{^{(1,2)}}}{^{\mu_2}}\nabla_{[\mu}\tensor{\zeta}{^{^{(3,3)}}}{_{\nu]\mu_1\mu_2}}$
      &
      $c_{2,0} = -\frac{1}{2}$
      \newline
      $c_{2,1} = -\frac{3}{2}$
      \newline
      $c_{2,2} = \frac{1}{2}$
      \\\hline
    \caption{Improvement terms for the shift symmetric vectors with $k=0,1,2$.}
    \label{tab:numerical2}
\end{longtable}

\begin{longtable}{|p{30pt}|p{30pt}|p{230pt}|p{150pt}|}
    \hline
      k & l & Improvement Terms & Constants\\\hline
      \multirow{2}{*}{0} 
      & 0 & 
      $\mathcal{I}_{(1,0)\mu\nu}^{(2,0)(0)}=\nabla_{[\mu}\tensor{\phi}{^{^{(2,0)}}}{_{\nu]}}{^{\mu_1}}\nabla_{\mu_1}\tensor{\zeta}{^{^{(0,2)}}}$ 
      \newline
      $\mathcal{I}_{(1,1)\mu\nu}^{(2,0)(0)}=\tensor{\phi}{^{^{(2,0)}}}{_{[\mu}}{^{\mu_1}}\nabla_{\nu]}\nabla_{\mu_1}\tensor{\zeta}{^{^{(0,2)}}}$ 
      &
      $c_{1,0} = -\frac{2}{3}$
      \newline
      $c_{1,1} = -2$
      \\\cline{2-4}
      & 1 &
      $\mathcal{I}_{(1,0)\mu\nu}^{(2,0)(1)}=\nabla_{[\mu}\tensor{\phi}{^{^{(2,0)}}}{_{\nu]}}{^{\mu_1}}\tensor{\zeta}{^{^{(1,2)}}}{_{\mu_1}}$ 
      \newline
      $\mathcal{I}_{(1,1)\mu\nu}^{(2,0)(1)}=\tensor{\phi}{^{^{(2,0)}}}{_{[\mu}}{^{\mu_1}}\nabla_{\nu]}\tensor{\zeta}{^{^{(1,2)}}}{_{\mu_1}}$ 
      \newline
      $\mathcal{I}_{(1,2)\mu\nu}^{(2,0)(1)}=\tensor{\phi}{^{^{(2,0)}}}{_{[\mu|}}{^{\mu_1}}\nabla_{\mu_1}\tensor{\zeta}{^{^{(1,2)}}}{_{|\nu]}}$ 
      &
      $c_{1,0} = -\frac{2}{3}$
      \newline
      $c_{1,1} = -\frac{4}{3}$
      \newline
      $c_{1,2} = -\frac{2}{3}$
      \\\cline{2-4}
      & 2 &
      $\mathcal{I}_{(1)\mu\nu}^{(2,0)(2)}=\tensor{\phi}{^{^{(2,0)}}}{_{[\mu}}{^{\mu_1}}\tensor{\zeta}{^{^{(2,2)}}}{_{\nu]\mu_1}}$
      &
      $c = -\frac{1}{3}$
      \newline
      $c_{1} = -\frac{4}{3}$
      \\\hline
      \multirow{3}{*}{1} 
      & 0 & 
      $\mathcal{I}_{(1,0)\mu\nu}^{(2,1)(0)}=\nabla_{[\mu}\tensor{\phi}{^{^{(2,1)}}}{_{\nu]}}{^{\mu_1}}\nabla_{\mu_1}\tensor{\zeta}{^{^{(0,3)}}}$ 
      \newline
      $\mathcal{I}_{(1,1)\mu\nu}^{(2,1)(0)}=\tensor{\phi}{^{^{(2,1)}}}{_{[\mu}}{^{\mu_1}}\nabla_{\nu]}\nabla_{\mu_1}\tensor{\zeta}{^{^{(0,3)}}}$ 
      \newline
      $\mathcal{I}_{(2,0)\mu\nu}^{(2,1)(0)}=\nabla^{\mu_1}\nabla_{[\mu}\tensor{\phi}{^{^{(2,1)}}}{_{\nu]}}{^{\mu_2}}\nabla_{\mu_1}\nabla_{\mu_2}\tensor{\zeta}{^{^{(0,3)}}}$ 
      \newline
      $\mathcal{I}_{(2,1)\mu\nu}^{(2,1)(0)}=\nabla^{\mu_1}\tensor{\phi}{^{^{(2,1)}}}{_{[\mu|}}{^{\mu_2}}\nabla_{\mu_1}\nabla_{|\nu]}\nabla_{\mu_2}\tensor{\zeta}{^{^{(0,3)}}}$ 
      &
      $c_{1,0} = (3+D)H^2$
      \newline
      $c_{1,1} = 2(1+D)H^2$
      \newline
      $c_{2,0} = -1$
      \newline
      $c_{2,1} = -2$
      \\\cline{2-4}
      & 1 &
      $\mathcal{I}_{(1,0)\mu\nu}^{(2,1)(1)}=\nabla_{[\mu}\tensor{\phi}{^{^{(2,1)}}}{_{\nu]}}{^{\mu_1}}\tensor{\zeta}{^{^{(1,3)}}}{_{\mu_1}}$ 
      \newline
      $\mathcal{I}_{(1,1)\mu\nu}^{(2,1)(1)}=\tensor{\phi}{^{^{(2,1)}}}{_{[\mu}}{^{\mu_1}}\nabla_{\nu]}\tensor{\zeta}{^{^{(1,3)}}}{_{\mu_1}}$ 
      \newline
      $\mathcal{I}_{(1,2)\mu\nu}^{(2,1)(1)}=\tensor{\phi}{^{^{(2,1)}}}{_{[\mu|}}{^{\mu_1}}\nabla_{\mu_1}\tensor{\zeta}{^{^{(1,3)}}}{_{|\nu]}}$
      \newline
      $\mathcal{I}_{(2,0)\mu\nu}^{(2,1)(1)}=\nabla^{\mu_1}\nabla_{[\mu}\tensor{\phi}{^{^{(2,1)}}}{_{\nu]}}{^{\mu_2}}\nabla_{\mu_1}\tensor{\zeta}{^{^{(1,3)}}}{_{\mu_2}}$ 
      \newline
      $\mathcal{I}_{(2,1)\mu\nu}^{(2,1)(1)}=\nabla^{\mu_1}\tensor{\phi}{^{^{(2,1)}}}{_{[\mu|}}{^{\mu_2}}\nabla_{\mu_1}\nabla_{|\nu]}\tensor{\zeta}{^{^{(1,3)}}}{_{\mu_2}}$ 
      \newline
      $\mathcal{I}_{(2,2)\mu\nu}^{(2,1)(1)}=\nabla^{\mu_1}\tensor{\phi}{^{^{(2,1)}}}{_{[\mu|}}{^{\mu_2}}\nabla_{\mu_1}\nabla_{\mu_2}\tensor{\zeta}{^{^{(1,3)}}}{_{|\nu]}}$
      \newline
      $\mathcal{I}_{(2,3)\mu\nu}^{(2,1)(1)}=\nabla^{\mu_1}\nabla_{[\mu}\tensor{\phi}{^{^{(2,1)}}}{_{\nu]}}{^{\mu_2}}\nabla_{\mu_2}\tensor{\zeta}{^{^{(1,3)}}}{_{\mu_1}}$
      \newline
      $\mathcal{I}_{(2,4)\mu\nu}^{(2,1)(1)}=\nabla^{\mu_1}\tensor{\phi}{^{^{(2,1)}}}{_{[\mu|}}{^{\mu_2}}\nabla_{\mu_2}\nabla_{|\nu]}\tensor{\zeta}{^{^{(1,3)}}}{_{\mu_1}}$ 
      &
      $c_{1,0} = \frac{1}{2}(3+2D)H^2$
      \newline
      $c_{1,1} = \frac{1}{6}(7+9D)H^2$
      \newline
      $c_{1,2} = \frac{1}{6}(5+3D)H^2$
      \newline
      $c_{2,0} = -\frac{2}{3}$
      \newline
      $c_{2,1} = -1$
      \newline
      $c_{2,2} = -\frac{1}{2}$
      \newline
      $c_{2,3} = -\frac{1}{3}$
      \newline
      $c_{2,4} = -\frac{1}{2}$
      \\\cline{2-4}
      & 2 &
      $\mathcal{I}_{(1)\mu\nu}^{(2,1)(2)}=\tensor{\phi}{^{^{(2,1)}}}{_{[\mu}}{^{\mu_1}}\tensor{\zeta}{^{^{(2,3)}}}{_{\nu]\mu_1}}$
      \newline
      $\mathcal{I}_{(2,0)\mu\nu}^{(2,1)(2)}=\nabla_{[\mu}\tensor{\phi}{^{^{(2,1)}}}{^{\mu_1\mu_2}}\nabla_{\nu]}\tensor{\zeta}{^{^{(2,3)}}}{_{\mu_1\mu_2}}$
      \newline
      $\mathcal{I}_{(2,1)\mu\nu}^{(2,1)(2)}=\nabla^{\mu_1}\tensor{\phi}{^{^{(2,1)}}}{_{[\mu}}{^{\mu_2}}\nabla_{\nu]}\tensor{\zeta}{^{^{(2,3)}}}{_{\mu_1\mu_2}}$
      \newline
      $\mathcal{I}_{(2,2)\mu\nu}^{(2,1)(2)}=\nabla_{[\mu|}\tensor{\phi}{^{^{(2,1)}}}{^{\mu_1\mu_2}}\nabla_{\mu_1}\tensor{\zeta}{^{^{(2,3)}}}{_{|\nu]\mu_2}}$
      \newline
      $\mathcal{I}_{(2,3)\mu\nu}^{(2,1)(2)}=\nabla^{\mu_1}\tensor{\phi}{^{^{(2,1)}}}{_{[\mu|}}{^{\mu_2}}\nabla_{\mu_2}\tensor{\zeta}{^{^{(2,3)}}}{_{|\nu]\mu_1}}$
      &
      $c = -\frac{1}{6}(3+D)H^2$
      \newline
      $c_{1} = -\frac{1}{3}(3+5D)H^2$
      \newline
      $c_{2,0} = -\frac{1}{3}$
      \newline
      $c_{2,1} = -\frac{2}{3}$
      \newline
      $c_{2,2} = -\frac{2}{3}$
      \newline
      $c_{2,3} = -\frac{1}{3}$
      \\\cline{2-4}
      & 3 &
      $\mathcal{I}_{(2,0)\mu\nu}^{(2,1)(3)}=\tensor{\phi}{^{^{(2,1)}}}{^{\mu_1\mu_2}}\nabla_{[\mu}\tensor{\zeta}{^{^{(3,3)}}}{_{\nu]\mu_1\mu_2}}$
      \newline
      $\mathcal{I}_{(2,1)\mu\nu}^{(2,1)(3)}=\nabla_{[\mu}\tensor{\phi}{^{^{(2,1)}}}{^{\mu_1\mu_2}}\tensor{\zeta}{^{^{(3,3)}}}{_{\nu]\mu_1\mu_2}}$
      \newline
      $\mathcal{I}_{(2,2)\mu\nu}^{(2,1)(3)}=\nabla^{\mu_1}\tensor{\phi}{^{^{(2,1)}}}{_{[\mu}}{^{\mu_2}}\tensor{\zeta}{^{^{(3,3)}}}{_{\nu]\mu_1\mu_2}}$
      &
      $c_{2,0} = \frac{1}{2}$
      \newline
      $c_{2,1} = -1$
      \newline
      $c_{2,2} = -1$
      \\\hline
    \caption{Improvement terms for the shift symmetric spin-2 field with $k=0,1$.}
    \label{tab:numerical3}
\end{longtable}

\vspace{-5pt}
\paragraph{Acknowledgments:}  KH acknowledges support from DOE grant DE-SC0009946.

\renewcommand{\em}{}
\bibliographystyle{utphys}
\addcontentsline{toc}{section}{References}
\bibliography{CKVhigherspin_arxiv_v2}

\providecommand{\href}[2]{#2}\begingroup\raggedright\begin{thebibliography}{10}

\bibitem{Noether:1918zz}
E.~Noether, ``{Invariant Variation Problems},''
  \href{http://dx.doi.org/10.1080/00411457108231446}{{\em Gott. Nachr.}
  {\bfseries 1918} (1918) 235--257},
  \href{http://arxiv.org/abs/physics/0503066}{{\ttfamily
  arXiv:physics/0503066}}.

\bibitem{Bonifacio:2018zex}
J.~Bonifacio, K.~Hinterbichler, A.~Joyce, and R.~A. Rosen, ``{Shift Symmetries
  in (Anti) de Sitter Space},''
  \href{http://dx.doi.org/10.1007/JHEP02(2019)178}{{\em JHEP} {\bfseries 02}
  (2019) 178}, \href{http://arxiv.org/abs/1812.08167}{{\ttfamily
  arXiv:1812.08167 [hep-th]}}.

\bibitem{Goon:2011qf}
G.~Goon, K.~Hinterbichler, and M.~Trodden, ``{Symmetries for Galileons and DBI
  scalars on curved space},''
  \href{http://dx.doi.org/10.1088/1475-7516/2011/07/017}{{\em JCAP} {\bfseries
  07} (2011) 017}, \href{http://arxiv.org/abs/1103.5745}{{\ttfamily
  arXiv:1103.5745 [hep-th]}}.

\bibitem{Goon:2011uw}
G.~Goon, K.~Hinterbichler, and M.~Trodden, ``{A New Class of Effective Field
  Theories from Embedded Branes},''
  \href{http://dx.doi.org/10.1103/PhysRevLett.106.231102}{{\em Phys. Rev.
  Lett.} {\bfseries 106} (2011) 231102},
  \href{http://arxiv.org/abs/1103.6029}{{\ttfamily arXiv:1103.6029 [hep-th]}}.

\bibitem{Burrage:2011bt}
C.~Burrage, C.~de~Rham, and L.~Heisenberg, ``{de Sitter Galileon},''
  \href{http://dx.doi.org/10.1088/1475-7516/2011/05/025}{{\em JCAP} {\bfseries
  05} (2011) 025}, \href{http://arxiv.org/abs/1104.0155}{{\ttfamily
  arXiv:1104.0155 [hep-th]}}.

\bibitem{Cheung:2014dqa}
C.~Cheung, K.~Kampf, J.~Novotny, and J.~Trnka, ``{Effective Field Theories from
  Soft Limits of Scattering Amplitudes},''
  \href{http://dx.doi.org/10.1103/PhysRevLett.114.221602}{{\em Phys. Rev.
  Lett.} {\bfseries 114} no.~22, (2015) 221602},
  \href{http://arxiv.org/abs/1412.4095}{{\ttfamily arXiv:1412.4095 [hep-th]}}.

\bibitem{Hinterbichler:2015pqa}
K.~Hinterbichler and A.~Joyce, ``{Hidden symmetry of the Galileon},''
  \href{http://dx.doi.org/10.1103/PhysRevD.92.023503}{{\em Phys. Rev. D}
  {\bfseries 92} no.~2, (2015) 023503},
  \href{http://arxiv.org/abs/1501.07600}{{\ttfamily arXiv:1501.07600
  [hep-th]}}.

\bibitem{Cheung:2016drk}
C.~Cheung, K.~Kampf, J.~Novotny, C.-H. Shen, and J.~Trnka, ``{A Periodic Table
  of Effective Field Theories},''
  \href{http://dx.doi.org/10.1007/JHEP02(2017)020}{{\em JHEP} {\bfseries 02}
  (2017) 020}, \href{http://arxiv.org/abs/1611.03137}{{\ttfamily
  arXiv:1611.03137 [hep-th]}}.

\bibitem{Novotny:2016jkh}
J.~Novotny, ``{Geometry of special Galileons},''
  \href{http://dx.doi.org/10.1103/PhysRevD.95.065019}{{\em Phys. Rev. D}
  {\bfseries 95} no.~6, (2017) 065019},
  \href{http://arxiv.org/abs/1612.01738}{{\ttfamily arXiv:1612.01738
  [hep-th]}}.

\bibitem{Padilla:2016mno}
A.~Padilla, D.~Stefanyszyn, and T.~Wilson, ``{Probing Scalar Effective Field
  Theories with the Soft Limits of Scattering Amplitudes},''
  \href{http://dx.doi.org/10.1007/JHEP04(2017)015}{{\em JHEP} {\bfseries 04}
  (2017) 015}, \href{http://arxiv.org/abs/1612.04283}{{\ttfamily
  arXiv:1612.04283 [hep-th]}}.

\bibitem{Bogers:2018zeg}
M.~P. Bogers and T.~Brauner, ``{Lie-algebraic classification of effective
  theories with enhanced soft limits},''
  \href{http://dx.doi.org/10.1007/JHEP05(2018)076}{{\em JHEP} {\bfseries 05}
  (2018) 076}, \href{http://arxiv.org/abs/1803.05359}{{\ttfamily
  arXiv:1803.05359 [hep-th]}}.

\bibitem{DeRham:2018axr}
C.~De~Rham, K.~Hinterbichler, and L.~A. Johnson, ``{On the (A)dS Decoupling
  Limits of Massive Gravity},''
  \href{http://dx.doi.org/10.1007/JHEP09(2018)154}{{\em JHEP} {\bfseries 09}
  (2018) 154}, \href{http://arxiv.org/abs/1807.08754}{{\ttfamily
  arXiv:1807.08754 [hep-th]}}.

\bibitem{Bonifacio:2019hrj}
J.~Bonifacio, K.~Hinterbichler, L.~A. Johnson, and A.~Joyce, ``{Shift-Symmetric
  Spin-1 Theories},'' \href{http://dx.doi.org/10.1007/JHEP09(2019)029}{{\em
  JHEP} {\bfseries 09} (2019) 029},
  \href{http://arxiv.org/abs/1906.10692}{{\ttfamily arXiv:1906.10692
  [hep-th]}}.

\bibitem{Bonifacio:2021mrf}
J.~Bonifacio, K.~Hinterbichler, A.~Joyce, and D.~Roest, ``{Exceptional scalar
  theories in de Sitter space},''
  \href{http://dx.doi.org/10.1007/JHEP04(2022)128}{{\em JHEP} {\bfseries 04}
  (2022) 128}, \href{http://arxiv.org/abs/2112.12151}{{\ttfamily
  arXiv:2112.12151 [hep-th]}}.

\bibitem{Dobrev:1977qv}
V.~K. Dobrev, G.~Mack, V.~B. Petkova, S.~G. Petrova, and I.~T. Todorov,
  \href{http://dx.doi.org/10.1007/BFb0009678}{{\em {Harmonic Analysis on the
  n-Dimensional Lorentz Group and Its Application to Conformal Quantum Field
  Theory}}}, vol.~63.
\newblock 1977.

\bibitem{Boers:2013pba}
M.~Boers, ``{Group theory and de Sitter QFT}: {The concept of mass},'' Master's
  thesis, Groningen U., 2013.

\bibitem{Basile:2016aen}
T.~Basile, X.~Bekaert, and N.~Boulanger, ``{Mixed-symmetry fields in de Sitter
  space: a group theoretical glance},''
  \href{http://dx.doi.org/10.1007/JHEP05(2017)081}{{\em JHEP} {\bfseries 05}
  (2017) 081}, \href{http://arxiv.org/abs/1612.08166}{{\ttfamily
  arXiv:1612.08166 [hep-th]}}.

\bibitem{Sun:2021thf}
Z.~Sun, ``{A note on the representations of SO(1,d + 1)},''
  \href{http://dx.doi.org/10.1142/S0129055X24300073}{{\em Rev. Math. Phys.}
  {\bfseries 37} no.~01, (2025) 2430007},
  \href{http://arxiv.org/abs/2111.04591}{{\ttfamily arXiv:2111.04591
  [hep-th]}}.

\bibitem{Sengor:2022lyv}
G.~Seng{\"o}r, ``{The de Sitter group and its presence at the late-time
  boundary},'' \href{http://dx.doi.org/10.22323/1.406.0356}{{\em PoS}
  {\bfseries CORFU2021} (2022) 356},
  \href{http://arxiv.org/abs/2206.04719}{{\ttfamily arXiv:2206.04719
  [hep-th]}}.

\bibitem{Sengor:2022kji}
G.~{\c{S}}eng{\"o}r, ``{Particles of a de Sitter Universe},''
  \href{http://dx.doi.org/10.3390/universe9020059}{{\em Universe} {\bfseries 9}
  no.~2, (2023) 59}, \href{http://arxiv.org/abs/2212.10626}{{\ttfamily
  arXiv:2212.10626 [hep-th]}}.

\bibitem{Enayati:2022hed}
M.~Enayati, J.-P. Gazeau, H.~Pejhan, and A.~Wang,
  \href{http://dx.doi.org/10.1007/978-3-031-16045-5}{{\em {The de Sitter (dS)
  Group and its Representations. An Introduction to Elementary Systems and
  Modeling the Dark Energy Universe}}}.
\newblock Synthesis Lectures on Mathematics {\&} Statistics. Springer, 2023.
\newblock \href{http://arxiv.org/abs/2201.11457}{{\ttfamily arXiv:2201.11457
  [math-ph]}}.

\bibitem{RiosFukelman:2023mgq}
A.~Rios~Fukelman, M.~Semp{\'e}, and G.~A. Silva, ``{Notes on gauge fields and
  discrete series representations in de Sitter spacetimes},''
  \href{http://dx.doi.org/10.1007/JHEP01(2024)011}{{\em JHEP} {\bfseries 01}
  (2024) 011}, \href{http://arxiv.org/abs/2310.14955}{{\ttfamily
  arXiv:2310.14955 [hep-th]}}.

\bibitem{Anninos:2023lin}
D.~Anninos, T.~Anous, B.~Pethybridge, and G.~{\c{S}}eng{\"o}r, ``{The discreet
  charm of the discrete series in dS$_{2}$},''
  \href{http://dx.doi.org/10.1088/1751-8121/ad14ad}{{\em J. Phys. A} {\bfseries
  57} no.~2, (2024) 025401}, \href{http://arxiv.org/abs/2307.15832}{{\ttfamily
  arXiv:2307.15832 [hep-th]}}.

\bibitem{Schaub:2024rnl}
V.~Schaub, ``{A Walk Through $Spin(1,d+1)$},''
  \href{http://arxiv.org/abs/2405.01659}{{\ttfamily arXiv:2405.01659
  [hep-th]}}.

\bibitem{Chen:2026kjo}
C.~Y.~R. Chen, L.~W. Lindwasser, and M.~Porrati, ``{Holomorphic structure of
  massive scalar fields in $\text{(A)dS}_2$},''
  \href{http://arxiv.org/abs/2601.20947}{{\ttfamily arXiv:2601.20947
  [hep-th]}}.

\bibitem{Hinterbichler:2024vyv}
K.~Hinterbichler, ``{Dualities among massive, partially massless and shift
  symmetric fields on (A)dS},''
  \href{http://dx.doi.org/10.1007/JHEP06(2024)033}{{\em JHEP} {\bfseries 06}
  (2024) 033}, \href{http://arxiv.org/abs/2402.16938}{{\ttfamily
  arXiv:2402.16938 [hep-th]}}.

\bibitem{Farnsworth:2024yeh}
K.~Farnsworth, K.~Hinterbichler, and S.~Saha, ``{Hidden conformal symmetry of
  the discrete series scalars in dS2},''
  \href{http://dx.doi.org/10.1103/PhysRevD.111.105002}{{\em Phys. Rev. D}
  {\bfseries 111} no.~10, (2025) 105002},
  \href{http://arxiv.org/abs/2410.19041}{{\ttfamily arXiv:2410.19041
  [hep-th]}}.

\bibitem{Lee:1999yu}
J.~Lee and S.~Lee, ``{Mass spectrum of D = 11 supergravity on AdS(2) x S**2 x
  T**7},'' \href{http://dx.doi.org/10.1016/S0550-3213(99)00598-2}{{\em Nucl.
  Phys. B} {\bfseries 563} (1999) 125--149},
  \href{http://arxiv.org/abs/hep-th/9906105}{{\ttfamily arXiv:hep-th/9906105}}.

\bibitem{Alkalaev:2019xuv}
K.~Alkalaev and X.~Bekaert, ``{Towards higher-spin AdS$_2$/CFT$_1$
  holography},'' \href{http://dx.doi.org/10.1007/JHEP04(2020)206}{{\em JHEP}
  {\bfseries 04} (2020) 206}, \href{http://arxiv.org/abs/1911.13212}{{\ttfamily
  arXiv:1911.13212 [hep-th]}}.

\bibitem{Conlon:2021cjk}
J.~P. Conlon, S.~Ning, and F.~Revello, ``{Exploring the holographic
  Swampland},'' \href{http://dx.doi.org/10.1007/JHEP04(2022)117}{{\em JHEP}
  {\bfseries 04} (2022) 117}, \href{http://arxiv.org/abs/2110.06245}{{\ttfamily
  arXiv:2110.06245 [hep-th]}}.

\bibitem{Apers:2022tfm}
F.~Apers, J.~P. Conlon, S.~Ning, and F.~Revello, ``{Integer conformal
  dimensions for type IIa flux vacua},''
  \href{http://dx.doi.org/10.1103/PhysRevD.105.106029}{{\em Phys. Rev. D}
  {\bfseries 105} no.~10, (2022) 106029},
  \href{http://arxiv.org/abs/2202.09330}{{\ttfamily arXiv:2202.09330
  [hep-th]}}.

\bibitem{Apers:2022vfp}
F.~Apers, ``{Aspects of AdS flux vacua with integer conformal dimensions},''
  \href{http://dx.doi.org/10.1007/JHEP05(2023)040}{{\em JHEP} {\bfseries 05}
  (2023) 040}, \href{http://arxiv.org/abs/2211.04187}{{\ttfamily
  arXiv:2211.04187 [hep-th]}}.

\bibitem{Plauschinn:2022ztd}
E.~Plauschinn, ``{Mass spectrum of type IIB flux compactifications
  {\textemdash} comments on AdS vacua and conformal dimensions},''
  \href{http://dx.doi.org/10.1007/JHEP02(2023)257}{{\em JHEP} {\bfseries 02}
  (2023) 257}, \href{http://arxiv.org/abs/2210.04528}{{\ttfamily
  arXiv:2210.04528 [hep-th]}}.

\bibitem{Blauvelt:2022wwa}
E.~Blauvelt, L.~Engelbrecht, and K.~Hinterbichler, ``{Shift Symmetries and
  AdS/CFT},'' \href{http://dx.doi.org/10.1007/JHEP07(2023)103}{{\em JHEP}
  {\bfseries 07} (2023) 103}, \href{http://arxiv.org/abs/2211.02055}{{\ttfamily
  arXiv:2211.02055 [hep-th]}}.

\bibitem{Arboleya:2025ocb}
{\'A}.~Arboleya, A.~Guarino, and M.~Morittu, ``{On type IIB AdS33 flux vacua
  with scale separation and integer conformal dimensions},''
  \href{http://dx.doi.org/10.22323/1.490.0184}{{\em PoS} {\bfseries CORFU2024}
  (2025) 184}, \href{http://arxiv.org/abs/2504.21508}{{\ttfamily
  arXiv:2504.21508 [hep-th]}}.

\bibitem{Bekaert:2025azj}
X.~Bekaert, A.~Sharapov, and E.~Skvortsov, ``{Higher-Spin Poisson Sigma Models
  and Holographic Duality for SYK Models},''
  \href{http://arxiv.org/abs/2509.19964}{{\ttfamily arXiv:2509.19964
  [hep-th]}}.

\bibitem{Brandt:1989gy}
F.~Brandt, N.~Dragon, and M.~Kreuzer, ``{Completeness and Nontriviality of the
  Solutions of the Consistency Conditions},''
  \href{http://dx.doi.org/10.1016/0550-3213(90)90037-E}{{\em Nucl. Phys. B}
  {\bfseries 332} (1990) 224--249}.

\bibitem{Wald:1990mme}
R.~M. Wald, ``{On identically closed forms locally constructed from a field},''
  \href{http://dx.doi.org/10.1063/1.528839}{{\em J. Math. Phys.} {\bfseries 31}
  no.~10, (1990) 2378}.

\bibitem{Dubois-Violette:1991dyw}
M.~Dubois-Violette, M.~Henneaux, M.~Talon, and C.-M. Viallet, ``{Some results
  on local cohomologies in field theory},''
  \href{http://dx.doi.org/10.1016/0370-2693(91)90527-W}{{\em Phys. Lett. B}
  {\bfseries 267} (1991) 81--87}.

\bibitem{Barnich:2000zw}
G.~Barnich, F.~Brandt, and M.~Henneaux, ``{Local BRST cohomology in gauge
  theories},'' \href{http://dx.doi.org/10.1016/S0370-1573(00)00049-1}{{\em
  Phys. Rept.} {\bfseries 338} (2000) 439--569},
  \href{http://arxiv.org/abs/hep-th/0002245}{{\ttfamily arXiv:hep-th/0002245}}.

\bibitem{Barnich:2001jy}
G.~Barnich and F.~Brandt, ``{Covariant theory of asymptotic symmetries,
  conservation laws and central charges},''
  \href{http://dx.doi.org/10.1016/S0550-3213(02)00251-1}{{\em Nucl. Phys. B}
  {\bfseries 633} (2002) 3--82},
  \href{http://arxiv.org/abs/hep-th/0111246}{{\ttfamily arXiv:hep-th/0111246}}.

\bibitem{Mikhailov:2002bp}
A.~Mikhailov, ``{Notes on higher spin symmetries},''
  \href{http://arxiv.org/abs/hep-th/0201019}{{\ttfamily arXiv:hep-th/0201019}}.

\bibitem{Maldacena:2011jn}
J.~Maldacena and A.~Zhiboedov, ``{Constraining Conformal Field Theories with A
  Higher Spin Symmetry},''
  \href{http://dx.doi.org/10.1088/1751-8113/46/21/214011}{{\em J. Phys. A}
  {\bfseries 46} (2013) 214011},
  \href{http://arxiv.org/abs/1112.1016}{{\ttfamily arXiv:1112.1016 [hep-th]}}.

\bibitem{Antunes:2025iaw}
A.~Antunes, N.~Levine, and M.~Meineri, ``{Demystifying integrable QFTs in AdS:
  No-go theorems for higher-spin charges},''
  \href{http://arxiv.org/abs/2502.06937}{{\ttfamily arXiv:2502.06937
  [hep-th]}}.

\bibitem{Eastwood:2002su}
M.~G. Eastwood, ``{Higher symmetries of the Laplacian},''
  \href{http://dx.doi.org/10.4007/annals.2005.161.1645}{{\em Annals Math.}
  {\bfseries 161} (2005) 1645--1665},
  \href{http://arxiv.org/abs/hep-th/0206233}{{\ttfamily arXiv:hep-th/0206233}}.

\bibitem{dairbekov2011conformalkillingsymmetrictensor}
N.~S. Dairbekov and V.~A. Sharafutdinov, ``On conformal killing symmetric
  tensor fields on riemannian manifolds,'' 2011.
\newblock \url{https://arxiv.org/abs/1103.3637}.

\bibitem{HEIL2016383}
K.~Heil, A.~Moroianu, and U.~Semmelmann, ``Killing and conformal killing
  tensors,''
  \href{http://dx.doi.org/https://doi.org/10.1016/j.geomphys.2016.04.014}{{\em
  Journal of Geometry and Physics} {\bfseries 106} (2016) 383--400}.
  \url{https://www.sciencedirect.com/science/article/pii/S0393044016301000}.

\bibitem{BRANSON1992314}
T.~P. Branson, ``Harmonic analysis in vector bundles associated to the rotation
  and spin groups,''
  \href{http://dx.doi.org/https://doi.org/10.1016/0022-1236(92)90050-S}{{\em
  Journal of Functional Analysis} {\bfseries 106} no.~2, (1992) 314--328}.
  \url{https://www.sciencedirect.com/science/article/pii/002212369290050S}.

\bibitem{Barnich:2005bn}
G.~Barnich, N.~Bouatta, and M.~Grigoriev, ``{Surface charges and dynamical
  Killing tensors for higher spin gauge fields in constant curvature spaces},''
  \href{http://dx.doi.org/10.1088/1126-6708/2005/10/010}{{\em JHEP} {\bfseries
  10} (2005) 010}, \href{http://arxiv.org/abs/hep-th/0507138}{{\ttfamily
  arXiv:hep-th/0507138}}.

\bibitem{Brust:2016gjy}
C.~Brust and K.~Hinterbichler, ``{Free {\ensuremath{\square}}$^{k}$ scalar
  conformal field theory},''
  \href{http://dx.doi.org/10.1007/JHEP02(2017)066}{{\em JHEP} {\bfseries 02}
  (2017) 066}, \href{http://arxiv.org/abs/1607.07439}{{\ttfamily
  arXiv:1607.07439 [hep-th]}}.

\bibitem{Hallowell:2005np}
K.~Hallowell and A.~Waldron, ``{Constant curvature algebras and higher spin
  action generating functions},''
  \href{http://dx.doi.org/10.1016/j.nuclphysb.2005.06.021}{{\em Nucl. Phys. B}
  {\bfseries 724} (2005) 453--486},
  \href{http://arxiv.org/abs/hep-th/0505255}{{\ttfamily arXiv:hep-th/0505255}}.

\bibitem{Hinterbichler:2022vcc}
K.~Hinterbichler, ``{Shift symmetries for p-forms and mixed symmetry fields on
  (A)dS},'' \href{http://dx.doi.org/10.1007/JHEP11(2022)015}{{\em JHEP}
  {\bfseries 11} (2022) 015}, \href{http://arxiv.org/abs/2207.03494}{{\ttfamily
  arXiv:2207.03494 [hep-th]}}.

\bibitem{Bonifacio:2023prb}
J.~Bonifacio and K.~Hinterbichler, ``{Fermionic shift symmetries in (anti) de
  Sitter space},'' \href{http://dx.doi.org/10.1007/JHEP04(2024)100}{{\em JHEP}
  {\bfseries 04} (2024) 100}, \href{http://arxiv.org/abs/2312.06743}{{\ttfamily
  arXiv:2312.06743 [hep-th]}}.

\bibitem{Hinterbichler:2022agn}
K.~Hinterbichler, D.~M. Hofman, A.~Joyce, and G.~Mathys, ``{Gravity as a
  gapless phase and biform symmetries},''
  \href{http://dx.doi.org/10.1007/JHEP02(2023)151}{{\em JHEP} {\bfseries 02}
  (2023) 151}, \href{http://arxiv.org/abs/2205.12272}{{\ttfamily
  arXiv:2205.12272 [hep-th]}}.

\end{thebibliography}\endgroup

\end{document}